\documentclass[12pt]{article}
\usepackage{eqsection,latexsym,epsf}
\usepackage{epsfig,amsfonts,cite}
\usepackage{amssymb}

\begin{document}

\newcommand{\<}{\langle}
\renewcommand{\>}{\rangle}
\def\us{\underline{\sigma}}
\def\qea{q_{\rm\scriptscriptstyle EA}}
\def\qeax{q^{\rm\scriptscriptstyle EA}_x}
\def\qei{q_{\rm\scriptscriptstyle EA}^{(i)}}
\def\qith{q_{\rm\scriptscriptstyle EA, th}^{(i)}}
\def\Rt{\widetilde{R}_{\rm th}}
\def\phg{\mbox{\boldmath $\phi$}}
\def\psg{\mbox{\boldmath $\psi$}}
\def\M{{\mathbb M}}
\def\hs{{\mbox{\boldmath {${\hat\sigma}$}}}}
\def\ga{{\cal G}_A}
\def\gb{{\cal G}_B}
\def\gc{{\cal G}_C}
\def\ha{{\cal H}_A}
\def\hb{{\cal H}_B}
\def\ua{\underline{\alpha}}

\title{Aging dynamics of heterogeneous spin models}

 \author{{ Andrea Montanari}              \\
  {\small\it Laboratoire de Physique Th\'{e}orique de l'Ecole Normale
  Sup\'{e}rieure\footnote {UMR 8549, Unit{\'e}   Mixte de Recherche du 
  Centre National de la Recherche Scientifique et de 
  l' Ecole Normale Sup{\'e}rieure. } }
   \\[-0.2cm]
   {\small\it 24, rue Lhomond, 75231 Paris CEDEX 05, FRANCE}        \\[-0.2cm]
   {\small Internet: {\tt Andrea.Montanari@lpt.ens.fr}}
           \\[0.5cm]
   { Federico Ricci-Tersenghi}              \\
   {\small\it Dipartimento di Fisica and SMC and UdR1 of INFM}\\[-0.2cm]
   {\small\it Universit\`a di Roma ``La Sapienza''}\\[-0.2cm]
   {\small\it Piazzale Aldo Moro 2, I-00185 Roma, ITALY}\\[-0.2cm]
   {\small Internet: {\tt Federico.Ricci@roma1.infn.it}}
           \\[-0.1cm]
   {\protect\makebox[5in]{\quad}}  
   \\
 }

\date{May 1, 2003}

\maketitle

\thispagestyle{empty}

\abstract{We investigate numerically the dynamics of
three different spin models in the aging regime. 
Each of these models is meant to be representative of 
a distinct class of aging behavior: coarsening systems,
discontinuous spin glasses, and continuous spin glasses.
In order to study dynamic heterogeneities
induced by quenched disorder, we consider
single-spin observables for a  given disorder realization.
In some simple cases we are 
able to provide analytical predictions for
single-spin response and correlation functions.

The results strongly depend upon the model considered.
It turns out that, by comparing the slow evolution of a few
different degrees of freedom, one can distinguish between 
different dynamic classes.  As a conclusion we present the
general properties which can be induced from our results, and
discuss their relation with thermometric arguments.}

\clearpage

\section{Introduction}

Physical systems with an extremely slow relaxation dynamics ({\it aging})
are, at the same time, ubiquitous and fascinating \cite{Struick}. 
Much of the insight we have on such systems comes
from the study of mean-field models \cite{DynamicsReview}.

One of the weak points of the results obtained so far is that they
focus on global quantities, e.g. the correlation and response
functions averaged over all the spins.  On the other hand, we expect
one of the peculiar features of glassy dynamics to be its {\it
heterogeneity} 
\cite{EdigerRev,Kegel,Weeks,Poole,Ediger,Kob,Bennemann,
RicciZecchinaLowT,BarratZecchinaLowT}.  
In order to understand this character, 
we study the out-of-equilibrium dynamics of three models
belonging to three different families 
of slowly evolving systems: coarsening systems, discontinuous 
and continuous glasses.

The correlation and response functions of a particular
spin depend (among the other things) upon its local environment,
i.e. the strength of its interactions with other spins. 
However, the way the single-spin dynamics is influenced by its 
environment is highly dependent upon the nature of the system 
as a whole. For instance, as we will show, while in coarsening systems
strongly interacting spins relax faster, for discontinuous 
glasses the opposite happens. Continuous glasses lie somehow
midway. In principle, this allows to distinguish different types 
of slow dynamics just by looking at the relation between a couple
of spins.

The local environment is not the only source of
heterogeneities \cite{CastilloEtAl1,CastilloEtAl2,CastilloEtAl3}.  
Even in systems without quenched disorder, 
like structural glasses, the thermal noise is able to break the
initial spatial uniformity and to bring the system in a strongly
heterogeneous configuration \cite{EdigerRev,Kegel,Weeks}. 
Nevertheless we think that considering systems with quenched
disorder can be an instructive first step even in that direction.
As it has been argued several times \cite{KirkpatrickThirumalai1,ModeCoupling},
structural glasses behaves similarly to some disordered systems because of
a sort of {\it self-induced} disorder. Each molecule relaxes 
in the amorphous environment produced by the (partially frozen)
arrangement of the other ones.
 
In order to extract quenched-disorder-induced heterogeneities
we will average on a very large number of independent thermal histories.  
This will delete the effects of the thermal noise.

On the contrary we are forced not to perform a naive average over the
disorder realizations, because this would wash any difference between
spins.  Instead of doing more careful disorder averages,
e.g.\ conditioning on the local environment of the spin under study, we
prefer to work with a unique fixed disorder realization.  In the limit
of large system size we expect local quantities still to fluctuate
from site to site, and to converge in distribution sense, making the
analysis of a single typical sample representative of the whole
{\it ensemble}.

After these preliminaries, we can summarize the approach used in
this work.  Given a disordered model, we take a
few typical samples\footnote{The complete
specification of the samples used in our simulations is available, 
upon request to one of the authors.} 
(as big as possible according to our numerical capabilities) 
from the ensemble and we repeat a huge number of times the
typical numerical experiment used for studying out-of-equilibrium
dynamics: start from a random configuration, quench the system to a
low temperature, where it evolves slowly, wait a time $t_w$, switch on
a small perturbing field, and take measurements.  The observables we
measure are local quantities, like single-spin correlation and
response functions, averaged over the thermal noise.

We shall consider three different disordered models:
\begin{itemize}
\item a two-dimensional ferromagnetic Ising model (couplings are all
ferromagnetic but of different strengths), which has a 
ferromagnetic phase below the critical temperature;
\item the 3-spin Ising model on random hypergraph,
which has a glassy phase with one step of replica symmetry breaking
(1RSB);
\item the  spin glass Ising model on random graph, also
known as Viana-Bray (VB) model \cite{VianaBray}, 
which is believed to have a glassy phase
with continuous replica symmetry breaking (FRSB).
\end{itemize}

The last two models are examples of {\it diluted mean-field}
spin glasses. They lack of any finite-dimensional geometric
structure: this makes them soluble using mean-field techniques.
On the other hand, the local fluctuations of quenched disorder
are not averaged out as in completely connected models.
For instance, the local connectivity is a Poissonian random
variable. Because of these two features, they are an interesting 
playing ground for understanding heterogeneous dynamics.

Diluted mean-field models have been intensively studied in
the last years, one of the qualifying motivations being their
correspondence with random combinatorial problems \cite{TCS}.  Statical
heterogeneities have been well understood, at least at 1RSB level. 
Throughout the paper we shall neglect
FRSB effects \cite{TwoStep}, and assume that 1RSB is a good approximation. 
In Refs. \cite{MarcGiorgioRiccardo,MarcRiccardo}, the authors defined a
linear-time algorithm that computes single-spin static quantities for a given
sample in 1RSB approximation. 
The algorithm was dubbed {\it surveys propagation} (SP) and,
strictly speaking, was defined for computing zero-temperature
quantities. It is straightforward,
although computationally more demanding,
to generalize it for finite temperatures $T$ (the generalization follows 
the ideas of Ref. \cite{MezardParisiBethe}): we shall
call this generalization SP${}_T$.

The resulting heterogeneities can be characterized
by a {\it local} Edwards-Anderson parameter. 
This can be defined  \cite{MonassonMarginal} by considering  
$m$ weakly-coupled 
``clones'' $\{\sigma^{(1)},\dots,\sigma^{(m)}\}$ of the system.
The local overlap between two of them $\qei(m) = 
\<\sigma_i^{(a)}\sigma_i^{(b)}\>$, with $a\neq b$, is given by
\begin{eqnarray}
\qei(m) \equiv \frac{1}{Z_m}\sum_{\alpha\in \{{\rm states}\}}\!\!
e^{-\beta m F_{\alpha}}\, \<\sigma_i\>_{\alpha}^2\, ,\label{LocalEA_Definition}
\end{eqnarray}
where the sum on $\alpha$ runs over the pure states,
$\<\cdot\>_{\alpha}$ denotes the thermal average over one of such states,
and $Z_m = \sum_\alpha e^{-\beta m F_{\alpha}}$.
Equation (\ref{LocalEA_Definition}) follows from the observation 
that the $m$ clones stay at any time in the same state $\alpha$, and that
each state is selected with probability $e^{-\beta m F_{\alpha}}/Z_m$. 
The  parameter $m$ enables us to select metastable states. 
In fact we expect the dynamics of discontinuous glasses to be
tightly related with the properties of high-energy metastable states
\cite{CugliandoloKurchanPspin,KurchanTAP}.

While in a paramagnetic phase $\qei(m)=0$ apart for a non-extensive 
subset of the spins, in the spin glass phase $\qei(m)>0$ in a finite 
fraction of the system. In general $\qei(m)$ depends upon the site $i$: 
the phase is heterogeneous.
We will return in the next Section on the dynamical significance of 
this and other statical results.

The paper is organized as follows. 
In Sec. \ref{GeneralitiesSection} we present some of the theoretical 
expectations which we are going to test. 
We also give a few technical details  concerning the numerics.
Section \ref{CoarseningSection} deals with coarsening systems.
We postulate the general behavior of response and correlation functions,
and test our predictions on a simple model. 
In Secs. \ref{DiscontinuousSection} and \ref{ContinuousSection} 
we present our numerical results for, respectively, the 3-spin 
and 2-spin interaction spin glasses on random (hyper)graphs. 
The particularly easy case of weakly interacting spins 
is treated in Sec. \ref{WeaklySection}. We show that the 
aging behavior of these spins can be computed from 
the behavior of their neighbors.
Finally, in Sec. \ref{DiscussionSection}, we discuss the general picture
which emerges from our observations. 
In Sec. \ref{ThermoApp} we interpret some of these properties
using thermometric arguments.
Appendix \ref{AppLargeN}
present some calculations for coarsening dynamics. 
A brief account of our results has appeared in Ref. 
\cite{Lettera}.
%
%
\section{Generalities}
\label{GeneralitiesSection}

In the following we shall discuss three different spin models.
Before embarking in such a tour it is worth presenting
the general frame and fixing some notations.

Our principal tools
will be the single-spin correlation and response functions:
\begin{eqnarray}
C_{ij}(t,t_w) \equiv \< \sigma_i(t)\sigma_j(t_w)\>\, ,\;\;\;\;\;\;\;
R_{ij}(t,t_w) \equiv 
\left.\frac{\partial  \<\sigma_i(t)\>}{\partial h_j(t_w)}\right|_{h=0}\, ,
\label{Functions}
\end{eqnarray}
where the average is taken with respect to some stochastic dynamics, and
$h_j$ is a magnetic field coupled to the spin $j$. 
It is also useful to define the integrated response 
$\chi_{ij}(t,t_w) = \int_{t_w}^t\!ds \, R_{ij}(t,s)$.

We shall not repeat the subscripts when considering 
the diagonal elements of the above functions
(i.e. we shall write $C_i$ for $C_{ii}$, etc.). 
The global (self-averaging) correlation and response functions are
obtained from the single-site quantities as follows:
$C(t,t_w) = (1/N)\sum_{i}C_{i}(t,t_w)$; $\chi(t,t_w) = 
(1/N)\sum_{i}\chi_{i}(t,t_w)$. The times $t$
and $t_w$ are measured with respect to the initial quench 
(at $t_{\rm quench}=0$) from infinite temperature.

We will be interested in comparing the outcome of static calculations
and out-of-equilibrium numerical simulations.
For instance, we expect the order parameter (\ref{LocalEA_Definition})
to have the following dynamical meaning\footnote{Notice that we shall always 
work with finite samples. The limits $\Delta t,t_w\to\infty$ 
must therefore be understood as $1\ll \Delta t,t_w \ll t_{\rm erg}(N)$. 
After a time of order $t_{\rm erg}(N)$ ergodicity is re-established 
and the system equilibrates.}
\begin{eqnarray}
\qei(m_{\rm th}) = \lim_{\Delta t\to\infty}
\lim_{t_w\to\infty}C_{i}(t_w+\Delta t,t_w)\, ,
\end{eqnarray}
where $m_{\rm th}$ is the parameter which select the highest 
energy metastable states.

In the aging regime $\Delta t,t_w\gg 1 $, $C_{i}(t_w+\Delta t,t_w)<\qei(m_{\rm th})$.
We expect the functions (\ref{Functions}) to satisfy the out-of-equilibrium
fluctuation-dissipation relation (OFDR) \cite{CugliandoloKurchanPspin}
\begin{eqnarray}
T R_{i}(t,t_w) = X_i[C_{i}(t,t_w)]\partial_{t_w}C_{i}(t,t_w) \label{OFDR}\, .
\end{eqnarray}
If $X_i[C_i]=1$ the fluctuation-dissipation theorem (FDT) is recovered. 
The arguments of Refs. \cite{StatDyn1,StatDyn2}, and the analogy with 
exactly soluble models 
\cite{CugliandoloKurchanPspin,CugliandoloKurchanSK,CugliandoloKurchanWeak}
suggest that the function $X_i[C]$ is related to the static
overlap probability distribution:
\begin{eqnarray}
P_i(q) = -\frac{dX_i(q)}{dq}\, .\label{StaticDynamic}
\end{eqnarray}
For discontinuous glasses the dynamics never approaches 
thermodynamically dominant
states. In this case the function $P_i(q)$ entering in Eq. 
(\ref{StaticDynamic}) is the overlap distribution among
highest metastable states.
We refer to the next Sections for concrete examples of
the general relation (\ref{StaticDynamic}).

Let us now give some details concerning our numerical simulations. 
We shall consider systems defined on $N$ Ising spins 
$\sigma_i=\pm 1$, $i\in\{1,\dots,N\}$,
with Hamiltonian $H(\sigma)$. The dynamics is defined by 
single spin flip moves with Metropolis acceptance rule.
The update will be {\it sequential} for the spin glass models 
of Secs. \ref{DiscontinuousSection} and \ref{ContinuousSection} and 
{\it random sequential} for the ferromagnetic model of 
Sec. \ref{CoarseningSection}.

For each one of the mentioned models, we shall repeat the typical aging 
``experiment''. The system is initialized in a random 
(infinite temperature) configuration. At time $t_{\rm quench}=0$, 
the system is cooled at temperature $T$ within its low temperature phase.
 We run the dynamics
for a ``physical'' time $t_w$ (corresponding to $t_w$ attempts to 
flip each spin). Then we ``turn on'' a small random magnetic field
$h_i=\pm h_0$ and go on running the Metropolis algorithm
for a maximum physical time  $\Delta t_{\rm MAX}$.
Notice that the random external field is changed at each 
trajectory.

The correlation and response of the single degrees of freedom are extracted
by measuring the following observables:
\begin{eqnarray}
C_{i}(t_w+2\Delta t,t_w|h_0) & \equiv &
\frac{1}{\Delta t} \sum_{t'=t_w+\Delta t+1}^{t_w+2\Delta t}
\<\sigma_i(t')\sigma_i(t_w)\>\, ,
\label{CiDefinition}\\
\chi_{i}(t_w+2\Delta t,t_w|h_0) & \equiv &
\frac{1}{\Delta t\, h_0} \sum_{t'=t_w+\Delta t+1}^{t_w+2\Delta t}
\<\sigma_i(t')\,{\rm sign}(h_i)\>\, ,
\label{ChiiDefinition}
\end{eqnarray}
where $\<\cdot\>$ denotes the average over the Metropolis trajectories and 
the random external field.
The sum over $t'$ has been introduced for reducing the statistical 
errors. While it 
is a drastic modification of the definition (\ref{Functions}) in the 
quasi-equilibrium regime $\Delta t\ll t_w$, it produces just a small 
correction in the aging regime $\Delta t,t_w\gg 1$. This correction should 
cancel out in two interesting cases: $(i)$ in the time sector
$t/t_w = {\rm const.}$, if one restrict himself to the
response-versus-correlation relation; $(ii)$ in ``slower'' time sectors
(e.g. $(\log t)^{\mu}-(\log t_w)^{\mu} = {\rm const.}$, with $\mu<1$).
The functions (\ref{CiDefinition}) and (\ref{ChiiDefinition}) have finite 
$h_0\to 0$ limits $C_i(t_w+\Delta t,t_w)$ and 
$\chi_i(t_w+\Delta t,t_w)$.

Finally, let us mention that we shall look at the
$\chi_i(t,t_w)$ versus $C_i(t,t_w)$ data from two different perspectives.
In the first one we focus on a fixed site $i$ and vary the times $t$ and 
$t_w$: this allows to verify the relations (\ref{OFDR}) and 
(\ref{StaticDynamic}). We shall refer to this type of 
presentations as {\it FD plots}. In the second approach we plot,
for a given couple of times, all the points $(C_i(t,t_w),\chi_i(t,t_w))$
for $i=1,\dots,N$.
Then we let $t$ grow as $t_w$ is kept fixed. We dubbed such a procedure
a {\it movie plot}. It emphasizes the relations between different 
degrees of freedom in the system. 
%
%
\section{Coarsening systems}
\label{CoarseningSection}

Coarsening is the simplest type of aging dynamics 
\cite{SpinodalReview,CoarseningReview}. Despite its simplicity it has 
many representatives: ferromagnets (both homogeneous and not), 
binary liquids, and, according  to the droplet model 
\cite{FisherHuse1,FisherHuse2,FisherHuse3,KoperHilhorst}, spin glasses.

Consider a homogeneous spin model with a low temperature
ferromagnetic phase (e.g. an Ising model in dimension $d\ge 2$).
When cooled below its critical temperature, the system 
quickly separates into domains of different magnetization. 
Within each domain the system is ``near'' one of its equilibrium 
pure phases. Nevertheless it keeps evolving at all times due the the 
growth of the domain size $\xi(t)$. This process is 
mainly driven by the energetics of domain boundaries.

In the $t\to\infty$ limit, the coarsening length obeys a power law
$\xi(t)\sim t^{1/z}$ (for non-conserved scalar order parameter
$z=2$). Two-times observables
decompose in a quasi-equilibrium part describing the fluctuations
within a domain ($C_{\rm eq}$ and $\chi_{\rm eq}$ in the equations below), 
plus an aging contribution which involves the motion  of the domain walls 
($C_{\rm ag}$, $C_{\rm dw}$ and $\chi_{\rm dw}$):
\begin{eqnarray}
C(t,t_w) & \approx & C_{\rm eq}(t-t_w) + 
\qea C_{\rm ag}\left(\frac{\xi(t)}{\xi(t_w)}\right) -
\qea t_w^{-a}C_{\rm dw}\left(\frac{\xi(t)}{\xi(t_w)}\right)\, ,
\label{Ferro_GlobalC}\\
\chi(t,t_w) & \approx & \chi_{\rm eq}(t-t_w) + 
\phantom{\qea C_{\rm ag}\left(\frac{\xi(t)}{\xi(t_w)}\right)}+
\qea t_w^{-a'} \chi_{\rm dw}\left(\frac{\xi(t)}{\xi(t_w)}\right)\, ,
\label{Ferro_GlobalChi}
\end{eqnarray}
where $q_{\rm EA}$ is the equilibrium Edwards-Anderson parameter.  For
a ferromagnet $q_{\rm EA} = M(\beta)^2$, $M(\beta)$ being the
spontaneous magnetization.  Moreover $C_{\rm eq}(\tau)$ decreases from
$(1-q_{\rm EA})$ to $0$ as $\tau$ goes from $0$ to $\infty$, and
$C_{\rm ag}(\lambda)$ goes from $1$ to $0$ as its argument increases from
$1$ to $\infty$.  Finally the equilibrium part of the susceptibility
$\chi_{\rm eq}(\tau)$ goes from $0$ to $(1-q_{\rm EA})/T$. In the case
of a scalar order parameter both the response and correlation
functions receive subleading contributions ($C_{\rm dw}$ and
$\chi_{\rm dw}$) from spins ``close'' to the domain walls. 
Notice that these spins will decorrelate faster and respond easier
than the others (in other words $C_{\rm dw}$ and $\chi_{\rm dw}$ are
typically positive). These contributions are expected to be 
proportional to the density of domain walls 
$\rho_{\rm dw}(t_w)\propto \xi(t_w)^{-1}$. This would imply $a = a' = 1/z$. 

It is easy to generalize this well-established scenario to include 
single-spin quantities in heterogeneous systems.
We expect that the quasi-equilibrium parts of the correlation 
and response functions will depend upon the detailed 
environment of each spin. On the other hand, the large scale motion 
of the domain boundaries will not depend upon the precise point of the
system we are looking at. Therefore the aging contribution will depend 
upon the site $x$ only through the local Edwards-Anderson parameter
$\qeax = M_x(\beta)^2$. 
The reason is that $\qeax$ quantifies the distance 
between pure phases as seen through the spin $\sigma_x$.

\begin{figure}
\centerline{\epsfig{figure=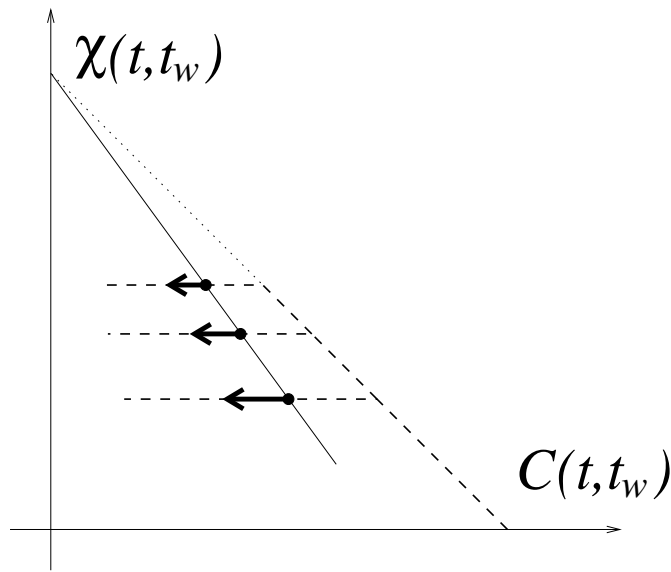,angle=0,width=0.5\linewidth}}
\caption{Qualitative picture of the response-versus-correlation 
plot for coarsening systems.
Bold dashed lines are the single-spin OFDR's. Black dots represent 
the correlation and response functions at a given pairs of times 
$(t,t_w)$. Arrows correspond to the ``velocities'' of the dots
when they move along the fluctuation-dissipation curves.}
\label{FerroGeneralPicture}
\end{figure}

We are led to propose the following form for single-site
functions:
\begin{eqnarray}
C_x(t,t_w) & \approx & C_x^{\rm eq}(t-t_w) + 
\qeax C_{\rm ag}\left(\frac{\xi(t)}{\xi(t_w)}\right) - 
\qeax t_w^{-a}C_{\rm dw}\left(\frac{\xi(t)}{\xi(t_w)}\right)\, ,
\label{CoarseningCorrelation}\\
\chi_x(t,t_w) & \approx & \chi_x^{\rm eq}(t-t_w) + 
\phantom{\qeax C_{\rm ag}\left(\frac{\xi(t)}{\xi(t_w)}\right)}+
\qeax t_w^{-a'} \chi_{\rm dw}\left(\frac{\xi(t)}{\xi(t_w)}\right)\, .
\label{CoarseningResponse}
\end{eqnarray}
This ansatz can be summarized, as far as $O(t_w^{-a},t_w^{-a'})$
terms are neglected, in the schematic response-versus-correlation
plot reported in Fig. \ref{FerroGeneralPicture}. 
Each spin follows its own fluctuation-dissipation
curve. This is composed by a quasi-equilibrium sector 
$T\chi_x = 1-C_x$, plus an horizontal aging sector $T\chi_x = 1-\qeax$.
Moreover, for each couple of times $t_w$ and $t$, all the points 
are aligned on the line passing through $(C=0,T\chi=1)$.

Notice that {\it spins with larger} $\qeax$ {\it relax faster}.
Roughly speaking this happens because they have to move a larger distance 
in order to jump from one pure phase to the other.

Due to the independence of $C_{\rm dw}$ and $\chi_{\rm dw}$ upon the
site $x$, Eqs. (\ref{CoarseningCorrelation}) and (\ref{CoarseningResponse})
imply that alignment in the $\chi$-$C$ plane is verified
even in the pre-asymptotic regime. This property is therefore more
robust than the OFDR which is violated by $O(t_w^{-a},t_w^{-a'})$ terms,
cf. Figs. \ref{Ferro_Type0Sites} and 
\ref{Ferro_FDMovie}.

Finally, both the large-$n$ calculation of Sec. \ref{LargeNSection},
and our numerical data, cf. Sec. \ref{NumericalCoarseningSection}, 
suggest that domain-wall contributions in Eqs. (\ref{CoarseningCorrelation}),
(\ref{CoarseningResponse}) have the same order of magnitude
$C_{\rm dw}(\lambda) \sim T \chi_{\rm dw}(\lambda)$.
%
%
\subsection{A staggered spin model}

Here we want to test our predictions in a simple context.
We shall consider a $d$-dimensional lattices ferromagnet, 
defined by the Hamiltonian
\begin{eqnarray}
H(\sigma) = -\sum_{(xy)}J_{xy}\sigma_x\sigma_y\, ,
\label{StaggeredModel}
\end{eqnarray}
where the sum runs over all the couples $(xy)$ of nearest-neighbors on the 
lattice $\mathbb{Z}^d$, and $J_{xy}\ge 0$. Moreover we assume periodicity
in the couplings. Namely, there exist positive integers
$l_1,\dots,l_d$ such that, for any $x,y\in \mathbb{Z}^d$, and
$\mu\in\{1,\dots,d\}$
\begin{eqnarray}
J_{xy} = J_{x+\hat{\mu}l_{\mu},y+\hat{\mu}l_{\mu}} 
\end{eqnarray}
where $\hat{\mu}$ is the unit vector in the $\mu$-th direction.
Clearly there are $V=l_1\cdot l_2\cdots l_d$ different 
``types'' of spins in this model. Two spins of the same type
have the same correlation and response functions. 
We can  identify these $V$ types with the spins
of the ``elementary cell'' $\Lambda\equiv \{x\in\mathbb{Z}^d| 0\le x_{\mu}<
l_{\mu}\}$.

Spatial periodicity is helpful  for two reasons:
$(i)$ it allows an analytical treatment in the large-$n$ limit; 
$(ii)$ averaging the single-spin quantities over the set of spins of 
a given type greatly improves the statistics of numerical simulations. 
%
%
\subsubsection{Large $n$}
\label{LargeNSection}

The model (\ref{StaggeredModel}) is easily generalized to 
$n$-vector spins $\phg_x = (\phi^1_x,\dots,\phi^n_x)$.
We just replace the ordinary product between spins in  Eq. 
(\ref{StaggeredModel}) with the scalar product.
Moreover we fix the spin length: $\phg_x\cdot\phg_x = n$. 
The dynamics is specified by the Langevin equation:
\begin{eqnarray}
\partial_t\phi^a_x(t) = -\zeta_x(t)\phi^a_x(t)+\sum_yJ_{xy}\phi^a_y(t)
+\eta^a_x(t)\, ,\label{Langevin}
\end{eqnarray}
where we introduced the Lagrange multipliers $\zeta_x(t)$ in order to enforce
the spherical constraint. The thermal noise is Gaussian with covariance
\begin{eqnarray}
\<\eta^a_x(t)\eta^b_y(s)\> = 2T\delta_{xy}\delta^{ab}\delta(t-s)\, .
\end{eqnarray}
The definition of correlation and response functions must be slightly
modified for an $n$-components order parameter:
\begin{eqnarray}
C_{xy}(t,t')  =  \frac{1}{n}\<\phg_x(t)\cdot \phg_y(t')\>\, ,\;\;\;\;\;\;\;\;
R_{xy}(t,t') = \frac{1}{n}\sum_a\frac{\delta\<\phi^a_x(t)\>}{\delta h_y^a(t')}\, .
\end{eqnarray}

Like its homogeneous relative \cite{CoarseningReview}, 
this model can be solved in the
limit $n\to\infty$. The calculations are outlined
in App. \ref{AppLargeN}. Let's summarize here the main results.
For $d> 2$ the model undergoes a phase transition at a finite temperature 
$T_c$. Below the critical temperature the $O(n)$ symmetry is broken: 
$\<\phi_x^a\>= M_x(\beta)\delta^{a1}$. Of course the spontaneous 
magnetizations preserves the spatial periodicity of the model:
$M_{x+l_{\mu}\hat{\mu}}(\beta) = M_x(\beta)$. 

\begin{figure}
\centerline{
\epsfig{figure=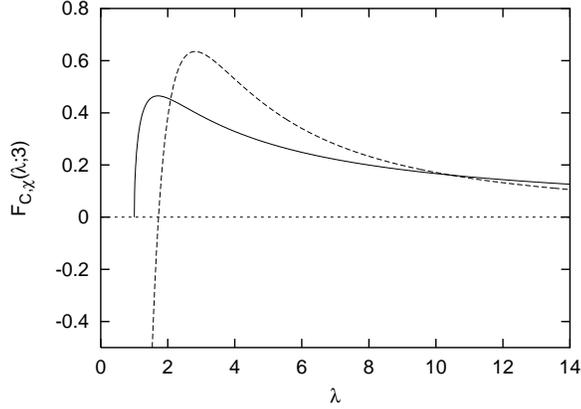,width=0.45\linewidth}}
\caption{Domain wall contributions to the correlation 
(dashed line) and integrated response functions (continuous line)
in the $n\to\infty$ limit: the universal scaling functions, 
cf. Eqs. (\ref{CoarseningCorrection1}) and 
(\ref{CoarseningCorrection2}), as a function of $\lambda = \sqrt{t/t_w}$.}
\label{CorrectionsFig}
\end{figure}
At low temperature, the forms 
(\ref{CoarseningCorrelation})-(\ref{CoarseningResponse}) hold, 
with $q_x^{\rm EA} = M^2_x(\beta)$,
$a = a' = d/2-1$, $z = 2$ and
\begin{eqnarray}
C_{\rm ag}(\lambda)  &= & \left(\frac{\lambda+\lambda^{-1}}{2}\right)^{-d/2}\, .\label{CoarseningLeading}
\end{eqnarray}
The subleading contribution read
\begin{eqnarray}
C_{\rm dw}(\lambda) & = & \frac{2T}{(8\pi)^{d/2}(\sum_{x\in\Lambda}M_x^2)
\Delta^{1/2}}\cdot {\cal F}_C(\lambda;d)\, ,\label{CoarseningCorrection1}\\
\chi_{\rm dw}(\lambda) & = & \frac{2}{(4\pi)^{d/2}(\sum_{x\in\Lambda}M_x^2)
\Delta^{1/2}}\cdot {\cal F}_{\chi}(\lambda;d)\, ,\label{CoarseningCorrection2}
\end{eqnarray}
where $\Delta$ is a constant which depends uniquely on the couplings $J_{xy}$,
cf. App. \ref{AppLargeN_Statics}.
${\cal F}_C(\cdot)$, ${\cal F}_{\chi}(\cdot)$ are two universal
function which do not depend either on the temperature or on the particular
model. The explicit expressions for these functions are 
not very illuminating. We report them  in App. 
\ref{AppLargeN_Dynamics}, see Eqs. (\ref{CcorrApp}) and (\ref{ChicorrApp}).
Here we plot the two functions in the $d=3$ case, 
see Fig. \ref{CorrectionsFig}.
Notice that both ${\cal F}_C(\lambda;d)$ and 
${\cal F}_{\chi}(\lambda;d)$ vanish in the $\lambda\to\infty$
limit. This could be expected because we know that $C_{x}(t,t_w)\to 0$ and
$\chi_{x}(t,t_w)\to (1-M_x^2)/T$ as $t\to\infty$ for any fixed $t_w$. 
%
%
\subsubsection{Numerical simulations}
\label{NumericalCoarseningSection}

\begin{figure}
\centerline{
\epsfig{figure=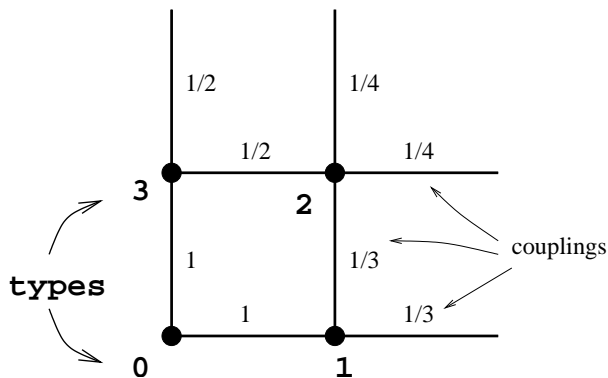,width=0.45\linewidth}}
\caption{Definition of the ferromagnetic couplings
for the two-dimensional model studied in 
Sec.~\ref{NumericalCoarseningSection}.}
\label{Types}
\end{figure}

We simulated the model (\ref{StaggeredModel}) in $d=2$
dimensions with $l_1=l_2=2$ and the choice of couplings among
spins in the elementary cell illustrated in Fig. \ref{Types}.
We used square lattices with linear size $L$.
There are $V=2^2$ different type of spins in this case. We 
shall improve our numerical estimates by averaging 
the single site functions $C_i(t,t_w;h_0)$ and 
$\chi_i(t,t_w;h_0)$, cf. Eqs. (\ref{CiDefinition}) and 
(\ref{CiDefinition}), over the $L^2/4$ spins of the same type.

Most our numerical results were obtained at temperature $T=1$.
A rough numerical estimate yields  $T_c = 1.10(5)$
for the critical temperature. The equilibrium magnetizations
for $T = 1$ of the four type of sites are:
$M_0 = 0.8803(5)$, $M_1 = 0.8395(5)$, $M_2 = 0.7573(5)$, and 
$M_3 = 0.8624(5)$. Notice that, in order to separate the magnetization
values on different sites, we are forced to choose a quite high temperature
for our simulations.

We expect the growth of the domain size in the model (\ref{StaggeredModel})
to follow asymptotically the law 
$\xi(t)\approx k(\beta)\cdot t^{1/z}$, with $z=2$, as in the 
homogeneous case. 
The pinning effect due to inhomogeneous couplings will renormalize
the coefficient $k(\beta)$. We checked this law by 
studying the evolution of the total magnetization 
starting from a random initial condition for different lattice sizes. 
It turns out that the law is reasonably well verified with a 
coefficient $k(\beta=1)$ of order one.

The aging ``experiment'' was repeated for several values of
the waiting time $t_w=10$, $10^2$, $10^3$, $10^4$, $10^5$. The correlation
and response functions were measured up to a maximum time 
interval (respectively)
$\Delta t_{\rm MAX} = 2^{10}$, $2^{13}$, $2^{15}$, $2^{17}$, $2^{19}$. 
The linear size of the
lattice was $L=2000$ in all the cases except for $t_w = 10^5$.
In this case we used $L=1000$. All the results were therefore
obtained in the $\xi(t)\ll L$ regime, with the exception,
possibly, of the latest times in the $t_w=10^5$ run. 
Some systematic discrepancies can be indeed noticed for these data.
In the Table below we report the number $N_{\rm stat}$ of different
runs for each choice of the parameters.

\vspace{0.4cm}
\begin{tabular}{|c||c|c|c|c|c|}
\hline
$h_0$  & $t_w=10$ & $t_w=10^2$ & $t_w=10^3$ & $t_w=10^4$  & $t_w = 10^5$\\
\hline
\hline
$0.025$ & $30$ & $23$ & $30$ & $9$ & $\star$ \\
\hline
$0.05$ & $30$ & $12$ & $12$ & $5$ & $10$\\
\hline
$0.10$ & $\star$ & $\star$ & $\star$ & $12$ & $\star$\\
\hline
\end{tabular}
\vspace{0.4cm}

\begin{figure}
\hspace{-1.5cm}
\begin{tabular}{cc}
\epsfig{figure=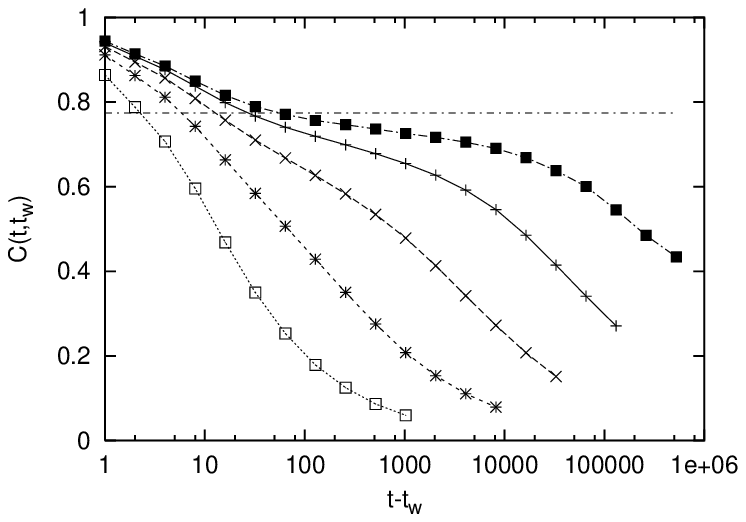,width=0.5\linewidth}&
\epsfig{figure=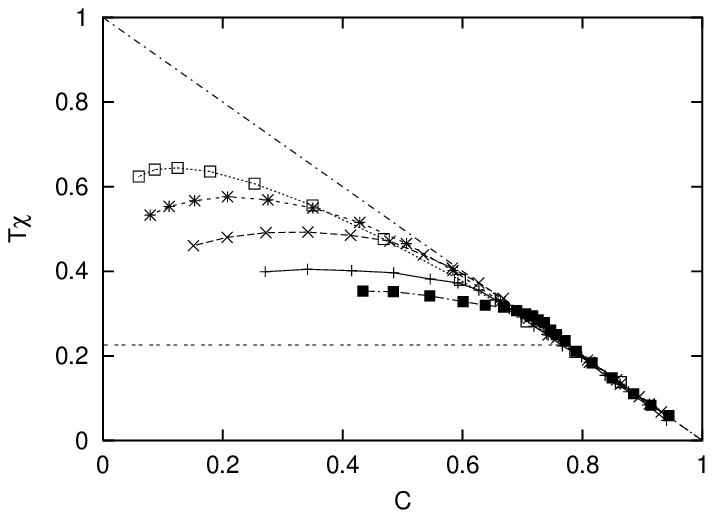,width=0.5\linewidth}
\end{tabular}
\caption{Correlation function and FD plot for type-0 sites (cf. Fig. 
\ref{Types}). Different symbols correspond to $t_w = 10$ ($\Box$),
$10^2$ ($\ast$), $10^3$ ($\times$), $10^4$ ($+$), $10^5$ (filled $\Box$). 
The dot-dashed line on the left is the equilibrium 
Edwards-Anderson parameter $M_0^2$. On the right we report the 
FDT line $T\chi = 1-C$ (dot-dashed) and the OFDR (dashed)
which corresponds to Eqs. (\ref{CoarseningCorrelation}) 
and (\ref{CoarseningResponse}).}
\label{Ferro_Type0Sites}
\end{figure}

Let us start by illustrating how the asymptotic behavior
summarized in Fig. \ref{FerroGeneralPicture} is approached.
In Fig. \ref{Ferro_Type0Sites}  we show the correlation 
functions and the FD plot
for type-0 sites. Notice that the approach to the asymptotic behavior
is quite slow
and, in particular, the domain-wall contribution to the response function
is pretty large. This can be an effect of the proximity of
the critical temperature: the ``thickness'' of the domain walls
grows with the equilibrium correlation length.
\begin{figure}
\hspace{-1.5cm}
\begin{tabular}{cc}
\epsfig{figure=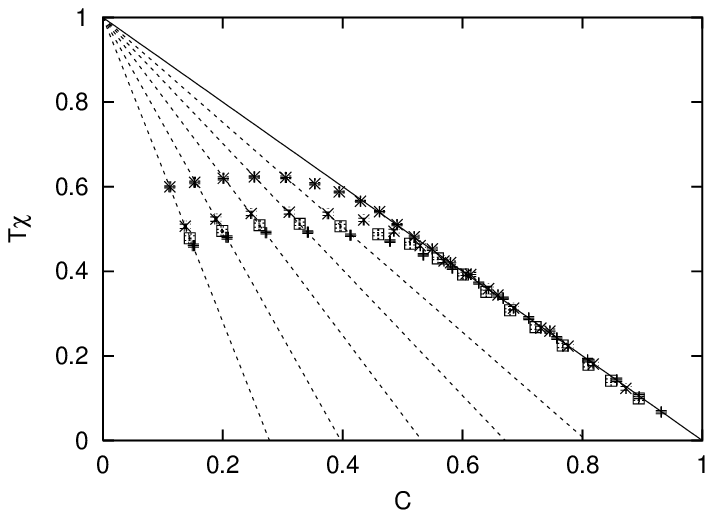,width=0.5\linewidth}&
\epsfig{figure=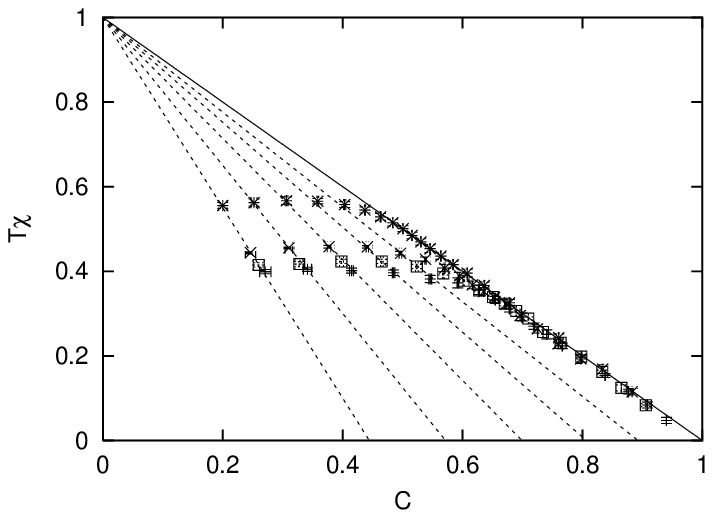,width=0.5\linewidth}
\end{tabular}
\caption{Movie plots at $t_w=10^3$ (left) and $10^4$ (right). The
various symbols correspond to different types of 
spin: type 0 ($+$), type 1 ($\times$), type 2 ($\Box$), and type 3 ($\ast$).
The straight lines confirm the alignment predicted in 
the general picture, cf. Fig. \ref{FerroGeneralPicture}.}
\label{Ferro_FDMovie}
\end{figure}
In Fig. \ref{Ferro_FDMovie} we verify the alignment of different 
sites correlation and response functions for a given pair of times
$(t,t_w)$. 
\begin{figure}
\begin{tabular}{cc}
\epsfig{figure=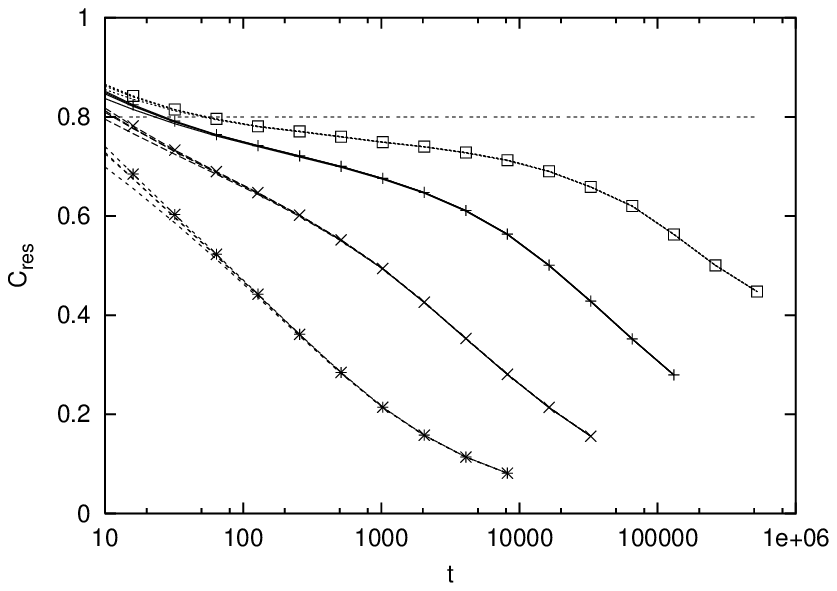,width=0.45\linewidth}&
\epsfig{figure=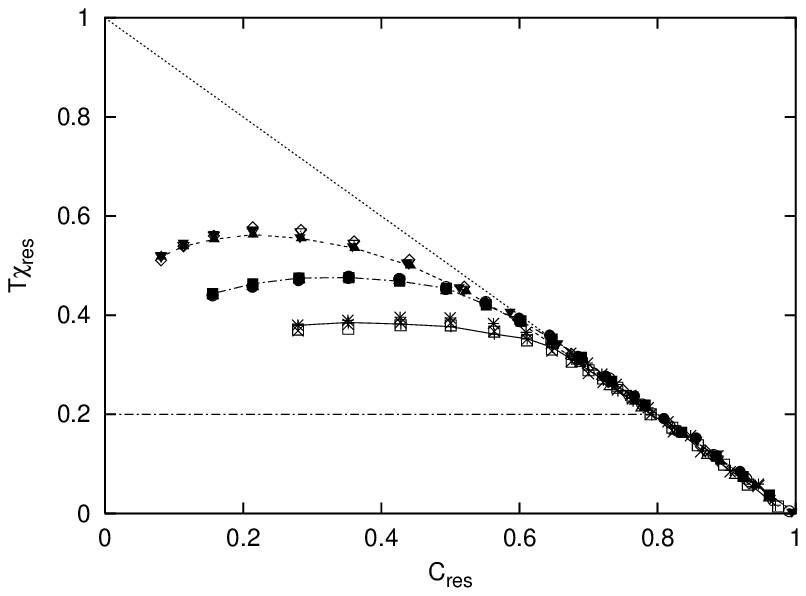,width=0.45\linewidth}
\end{tabular}
\caption{Correlation function (left) and FD plot (right) with the 
rescaled correlation and response functions,
see Eq. (\ref{FerroRescaled}), for all the four spin types and
several different waiting times:  $t_w =10^2$, $10^3$, $10^4$, and, on
the left $10^5$. Here $h_0 = 0.05$.}
\label{Ferro_FDres}
\end{figure}
Notice that the alignment works quite well even for 
``pre-asymptotic'' times, i.e. when the anomalous response
is still sizeable and the 
OFDR is not well verified, cf. Fig. \ref{Ferro_Type0Sites}.

\begin{figure}
\centerline{
\epsfig{figure=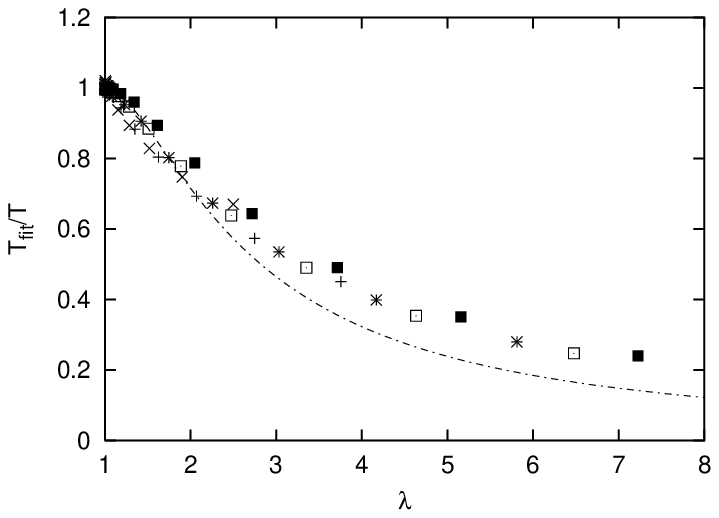,width=0.6\linewidth}}
\caption{The fitting temperature (\ref{FittingTemperature}) as a function
of $\lambda=\sqrt{t/t_w}$ for $t_w=10$ (filled $\Box$), 
$10^2$ ($\Box$), $10^3$ ($\ast$), $10^4$ ($+$), and $10^5$ ($\times$).
The dot-dashed line is the  $n=\infty$ scaling function 
(\ref{CoarseningLeading}), with $d=2$.}
\label{Ferro_Slope}
\end{figure}
In order to check the form (\ref{CoarseningResponse}) for the site 
dependence of the domain wall contribution, we plot in Fig. 
\ref{Ferro_FDres} the rescaled response and correlation functions:
\begin{eqnarray}
C^{\rm res}_x = \frac{\overline{q}}{\qeax}C_x\, ,\;\;\; 
T\chi^{\rm res}_x = 1-\frac{\overline{q}}{\qeax}(1-T\chi_x)\, ,
\label{FerroRescaled}
\end{eqnarray}
where $\overline{q}$ is an arbitrary reference overlap. The rescaled
correlation and response functions of different types of spin
coincide perfectly for any couple of times $(t,t_w)$.

Finally, we notice that we can consistently define
a time-dependent {\it fitting temperature} as the slope of the lines
in Fig. \ref{Ferro_FDMovie}, i.e.:
\begin{eqnarray}
T_{\rm fit}(t,t_w) = \frac{T\, C_x(t,t_w)}{1-T\chi_x(t,t_w)}\, .
\label{FittingTemperature}
\end{eqnarray}
As a consequence of Eqs. (\ref{CoarseningCorrelation}) 
and (\ref{CoarseningResponse})
this temperature should depend upon $t$ and $t_w$ only
through the parameter $\lambda = \xi(t)/\xi(t_w)$. In Fig. \ref{Ferro_Slope}
we verify this scaling.
%
%
\section{Discontinuous glasses}
\label{DiscontinuousSection}

In this Section we consider a ferromagnetic Ising model with 3-spin 
interactions, defined on a random {\it hypergraph} \cite{XorSat,Creignou}. 
More precisely, the Hamiltonian reads
\begin{eqnarray}
H(\sigma) = -\sum_{(ijk)\in{\cal H}} \sigma_i\sigma_j\sigma_k\, .
\label{ThreeSpinHamiltonian}
\end{eqnarray}
The hypergraph ${\cal H}$ defines which triplets of spins 
do interact. We construct it by randomly choosing $M$ among the 
$N(N-1)(N-2)/3!$ possible triplets of spins.

Although ferromagnetic, this model is thought to have a glassy behavior, 
due to {\it self-induced} frustration \cite{FranzEtAlFerromagn}.
Depending upon the value of $\gamma\equiv M/N$, it undergoes 
no phase transition (if $\gamma<\gamma_d$), a purely dynamic
phase transition  (if $\gamma_d<\gamma<\gamma_c$), or a dynamic and a 
static phase transitions (if $\gamma>\gamma_c$) as the temperature 
is lowered. The 1RSB analysis of
Refs. \cite{XorSat,FranzEtAlExact} yields 
$\gamma_d\approx 0.818$ and $\gamma_c\approx 0.918$.
These results have been later confirmed by rigorous derivations
\cite{CoccoEtAlXorSat,MezardEtAlXorSat}.

We studied two samples extracted from the {\it ensemble} defined
above: the first one involves $N=100$ sites and $M=100$ interactions
(hereafter we shall refer to it as $\ha$); 
in the second one ($\hb$) we have $N=M=1000$. In both cases  
$\gamma=1>\gamma_c$. 
The hypergraph $\ha$ consists of a large connected component 
including $96$ sites, plus $4$ isolated sites 
(namely the sites $i=15,22,62,69$). The largest connected component
of $\hb$ includes $938$ sites (there are $62$ isolated sites). 
We will illustrate our results mainly on
$\ha$ (on this sample we were able to reach larger waiting times). 
$\hb$ has been used to check finite-size effects.

\begin{figure}
\begin{tabular}{cc}
\epsfig{figure=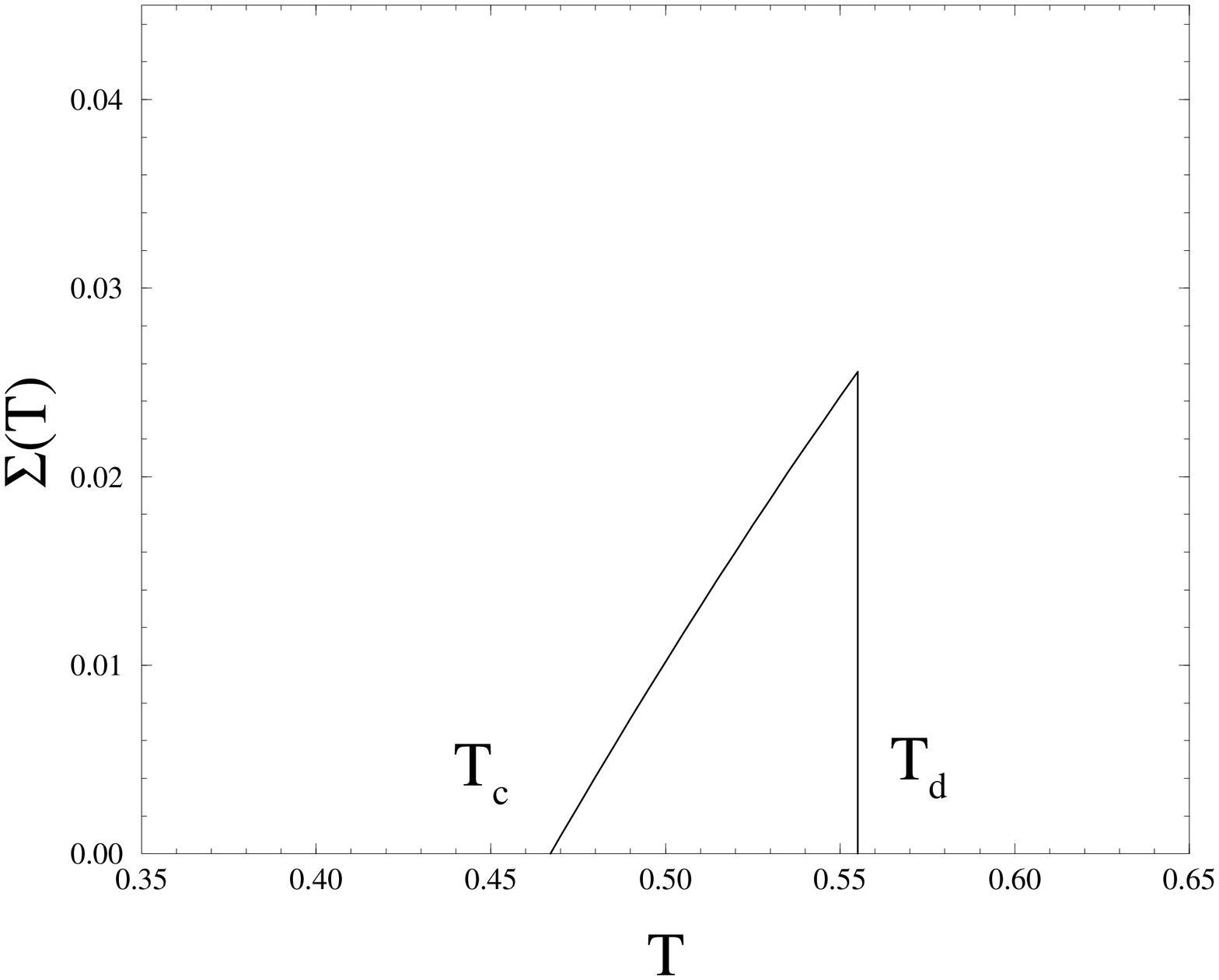,angle=-90,width=0.45\linewidth}&
\epsfig{figure=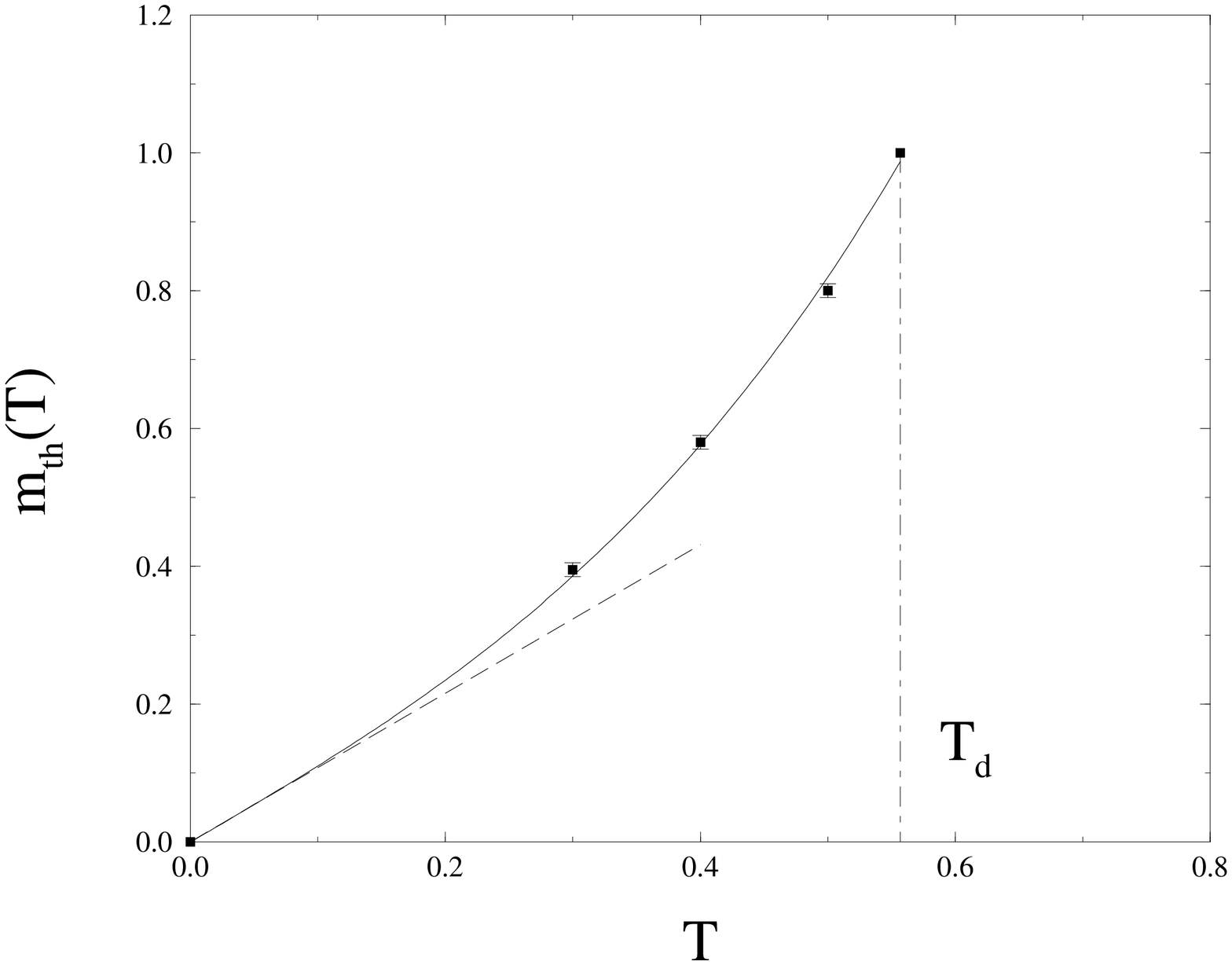,angle=-90,width=0.45\linewidth}
\end{tabular}
\caption{The complexity $\Sigma(T)$ (left) and the 1RSB parameter 
$m_{\rm th}(T)$ (right) for threshold states as functions of the 
temperature $T$. These curves refer to  sample $\ha$ considered in 
Sec.~\ref{DiscontinuousSection}.}
\label{cplxT}
\end{figure}
Using SP${}_T$, we computed the 
1RSB free energy density $F(m,\beta)$ and complexity 
$\Sigma(T) = \beta\,\partial_m F(m,\beta)|_{m=1}$ 
for our samples as a function of the temperature $T=1/\beta$. 
The resulting complexity is reported in Fig. \ref{cplxT} for
sample $\ha$. The 
dynamic and static temperatures are defined, respectively, as the points
where a non-trivial (1RSB) solution of the cavity equations first appears, and
where its complexity vanishes. From the results of Fig. \ref{cplxT} 
(left frame) we get the estimates $T_d = 0.557(2)$ and $T_c = 0.467(2)$.

In analogy with the analytic solution of the $p$-spin spherical model 
\cite{CugliandoloKurchanPspin,KurchanTAP}, we assume the aging 
dynamics of the model (\ref{ThreeSpinHamiltonian}) to be dominated 
by threshold states. These are defined as the 1RSB metastable 
states with the highest free energy density.
Although not exact \cite{TwoStep}, we expect this assumption to be a good
approximation for not-too-high values of $\gamma$.
The threshold 1RSB parameter $m_{\rm th}(T)$
can be computed by imposing the condition $\partial_m^2[m F(m,\beta)]=0$.
We computed $m_{\rm th}(T)$ on  sample $\ha$ for a few temperatures 
below $T_d$. We get $m_{\rm th}(0.3)=0.395(10)$, 
$m_{\rm th}(0.4)=0.58(1)$, $m_{\rm th}(0.5)=0.80(1)$.
Moreover, in the zero temperature limit, we obtain 
$m_{\rm th}(T) =\mu_{\rm th}T+O(T^2)$,
with $\mu_{\rm th} =1.08(1)$. These results are summarized in Fig. \ref{cplxT} 
(right frame). A good description of the temperature dependence 
is obtained using the polynomial fit $m_{\rm th}(T) 
= 1.08\, T+0.038\, T^2+2.17\, T^3$
(cf. continuous line in Fig. \ref{cplxT}, right frame).

Now we are in the position of precising the connection between 
single-spin statics and aging dynamics, outlined in Sec.
\ref{GeneralitiesSection}. It is convenient to work with 
the integrated response functions $\chi_i(t,t_w)$.
Equation (\ref{OFDR}) implies the relation
$\chi_i(t,t_w) = \chi_i[C_i(t,t_w)]$ to hold in the limit 
$t,t_w\to\infty$. 
Within a 1RSB approximation, Eq. (\ref{StaticDynamic}) corresponds to:
\begin{eqnarray}
T\chi_i[q] = \left\{\begin{array}{cl}
1-q&\mbox{ for $q>\qith$}\, ,\\
1-\qith-m_{\rm th}(q-\qith)&\mbox{ for $q<\qith$}\, ,\end{array}
\right.\label{OneStepChi}
\end{eqnarray}
where we used the shorthand $\qith = \qei(m_{\rm th})$. Since
the SP${}_T$ algorithm allows us to compute both $m_{\rm th}$ and
the parameters $\qei(m)$ for a given sample in linear time,
we can check the above prediction in our simulations.
%
%
\subsection{Numerical results}
\label{DiscontinuousNumericalSection}

We ran our simulations at three different temperatures ($T=0.3,0.4,0.5$)
and intensities of the external field ($h_0=0.05,0.1,0.15$).
In order to probe the aging regime, we repeated our simulations for 
several waiting times $t_w=10, 10^2, 10^3, 10^4$, with
(respectively) $\Delta t_{\rm MAX} = 2^{13}, 2^{16}, 2^{16},2^{18}$.

We summarize in the Table below the statistic of our simulations on sample
$\ha$. 

\vspace{0.3cm}
\begin{tabular}{|c||c|c|c|c|}
\hline
$h_0$  & $t_w=10$ & $t_w=10^2$ & $t_w=10^3$ & $t_w=10^4$  \\
\hline
\hline
$0.05$ & $5\cdot 10^6$ & $5\cdot 10^6$ & $5\cdot 10^6$ & $10^6$ \\
\hline
$0.10$ & $1.5\cdot 10^6$ & $1.5\cdot 10^6$ & $1.5\cdot 10^6$ & $10^6$ \\
\hline
$0.15$ & $10^6$ & $10^6$ & $10^6$ & $0.5\cdot 10^6$ \\
\hline
\end{tabular}
\vspace{0.3cm}

For sample $\hb$, we limited ourselves to the case $h_0=0.10$, $T=0.4$,
and generated $0.9\cdot 10^6$ Metropolis trajectories with 
$t_w=10^3$. 
%
%
\subsubsection{Two types of spins}
\label{TwoTypesSection}

The most evident feature of our numerical data, is that the spins 
can be clearly
classified in two groups: (I) the ones which behave as if the system 
were in equilibrium: the corresponding correlation and response functions
satisfy time-traslation invariance and FDT; 
(II) the out-of-equilibrium spins, whose correlation and response
functions are non-homogeneous on long time scales and violate
FDT. 

Of course the group (I) includes the isolated sites, but
also an extensive fraction of  non-isolated sites 
(for instance  the $12$ sites $i=1,6,8,14,27,39,68,74,$
$77,84,87,98$ of sample $\ha$). 
Remarkably these sites are the ones for which the 
SP${}_T$ algorithm 
returns $q^{(i)}_{\rm EA}=0$: they are paramagnetic from the static  
point of view.
\begin{figure}
\begin{tabular}{cc}
\hspace{-0.25cm}\epsfig{figure=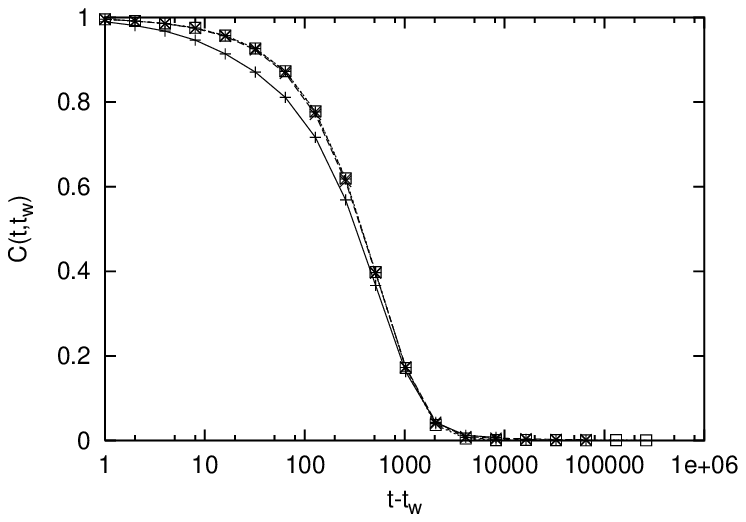,width=0.475\linewidth}&
\hspace{-0.5cm}\epsfig{figure=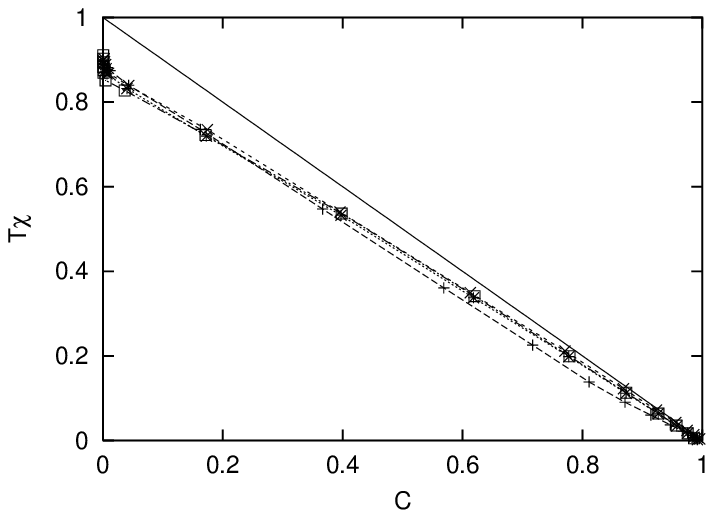,width=0.475\linewidth}
\end{tabular}
\caption{Correlation function (left) and  
FD plot (right) of the spin $i=1$ (sample $\ha$) for 
$T=0.5$, $h_0=0.1$ and $t_w=10 \div 10^4$. Time-traslation
invariance is well verified for $t_w\gtrsim 100$. The discrepancy
from FDT (continuous line on the right) can 
be ascribed to nonlinear response effects.}
\label{Site1_example}
\end{figure}
\begin{figure}
\begin{tabular}{cc}
\hspace{-0.25cm}\epsfig{figure=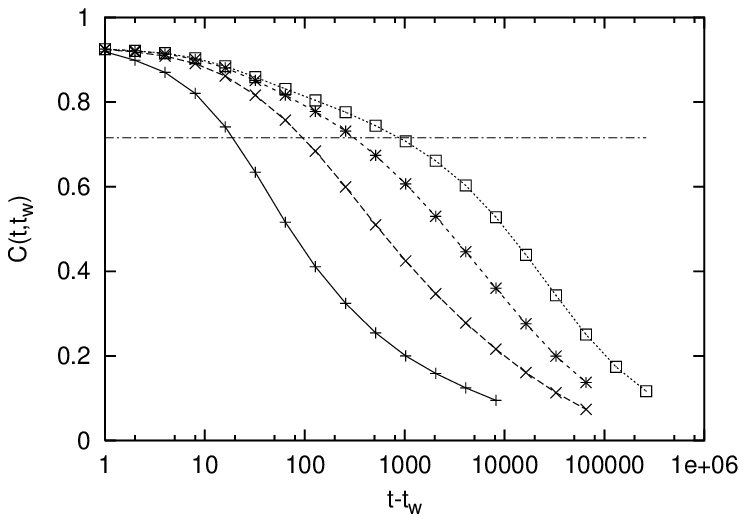,width=0.475\linewidth}&
\hspace{-0.5cm}\epsfig{figure=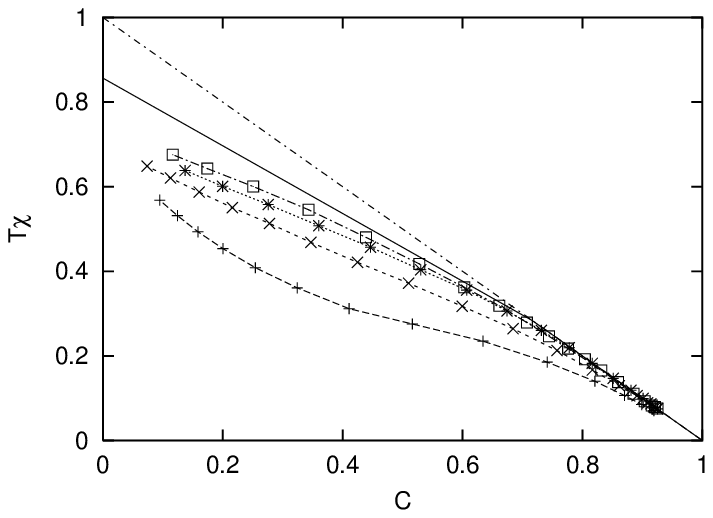,width=0.475\linewidth}
\end{tabular}
\caption{Correlation function (left) and FD plot (right)
for the spin $i=0$ (sample $\ha$).  The dashed line on the left 
corresponds the ergodicity breaking parameter 
$\qei=0.716(7)$ obtained with the SP${}_T$ algorithm. 
On the right we report with a full line the corresponding OFDR 
predicted within a 1RSB scenario.}
\label{Site0_example}
\end{figure}

In Figs. \ref{Site1_example} and \ref{Site0_example} we present
the correlation function and the FD plot, respectively,
for a type-I site and a type-II site. In both cases we took 
$T=0.5$ and $h=0.05$.

There exists a nice geometrical characterization of type-I 
sites in terms of a {\it leaf-removal algorithm} 
\cite{CoccoEtAlXorSat,MezardEtAlXorSat}. Let us recall
here the definition of this procedure. 
The algorithm starts by removing  all the interactions which involve at 
least one site with connectivity 1.
The same operation is repeated recursively until no connectivity-1
site is left. The reduced graph will contain either
isolated sites, or sites which have connectivity greater than one.
The sites of this last type are surely type-II, but they are not the only ones.
In fact one has to restore a subset of the original
interactions according to the following recursive rule.
If an interaction involves at least two type-II sites,
restore it and declare the third site to be type-II. 
If no such an interaction can be found among the original ones, stop.
In this way, one has singled out the subset of original interactions
which are {\it relevant} for aging dynamics.
The sites which remain isolated after this 
restoration procedure are type-I sites, the others are type-II. 

The dynamical relevance of this construction is easily understood
by considering two simple cases. A connectivity-1 spin whose two
neighbors evolve slowly will be affected by a slow local field and 
will relax on the same time-scale of the field. In the opposite case,
two connectivity-1 spins whose neighbor evolve slowly will
effectively see just a slowly alternating two-spin coupling between them.
They will relax as fast as a two-spin isolated cluster does.

\begin{figure}
\centerline{
\epsfig{figure=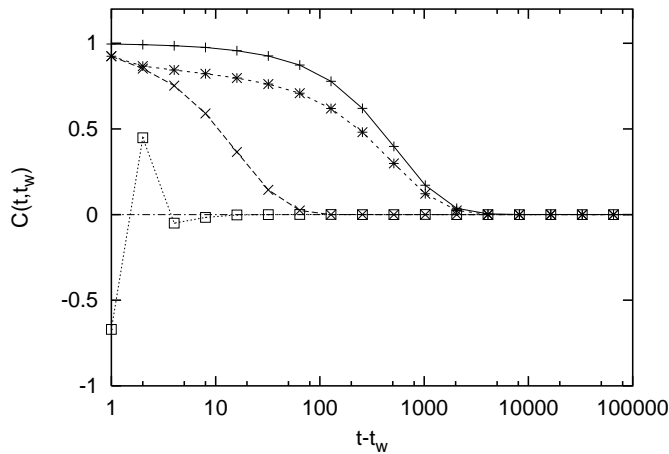,width=0.5\linewidth}}
\caption{The correlation functions of four type-I spins of sample 
$\ha$: from above to below $i=1,14,8,15$. The first three sites are connected 
to the rest of the cluster (and therefore interacting), the last one 
is isolated (free).}
\label{FDTSites}
\end{figure}
It is worth stressing that the above construction does not 
contain all the dynamical  information on the model.
For instance, one may wonder whether the dynamics of type-I spins does 
resemble to the dynamics of isolated spins. The answer is given
in Fig. \ref{FDTSites} where we reproduce the correlation functions
for several different type-I spins, for $T=0.5$, $h=0.1$ and $t_w=10^4$
(remember that the dependence upon $t_w$ is weak for these sites).
The results are strongly  site-dependent and 
by no way similar to the free-spin case. 
Notice the peculiar behavior of the isolated spin,
an artifact of Metropolis algorithm with sequential updatings.
Weren't for the perturbing field we would have $\sigma_i(t) = 
(-1)^t\sigma_i(0)$, which implies $C_i(t_w+1,t_w)=-1$,
$C_i(t_w+2,t_w)=1$, and $C_i(t_w+2\Delta t,t_w)=0$ for $\Delta t\ge2$
[remind the time average in Eq.(\ref{CiDefinition})].
In the presence of an external field the correlation function (with no
time-average) becomes $C_i(t_w+t,t_w)=[-\exp(-2\beta h)]^t$.

Finally, in Fig. \ref{Site0_example} we compare the numerical results
with the prediction from the statics, cf. Eq. (\ref{OneStepChi}).
The agreement is quite good although finite-$t_w$ and
finite-$h_0$ effects are not negligible.
%
%
\subsubsection{Dependence on the perturbing field}
\label{FieldSection}

We claimed that type-I spins satisfy FDT. However the FD plot in 
Fig. \ref{Site1_example} is not completely convincing on this point.
The data show a small but significant discrepancy with respect 
to the FDT-line. We want to show here that this is (at least partially)
a finite-$h_0$ effect.

In order to have a careful check of this effect, we used, on sample $\ha$, 
three different values of $h_0$ for any choice of $T$ and $t_w$.
The main outcome of this analysis is that the $h_0$ dependence
is much stronger for type-I than for type-II spins. This can be 
easily understood: type-II spins interact strongly with their 
neighbors and are therefore ``stiffer''.

\begin{figure}
\begin{tabular}{cc}
\hspace{-0.25cm}\epsfig{figure=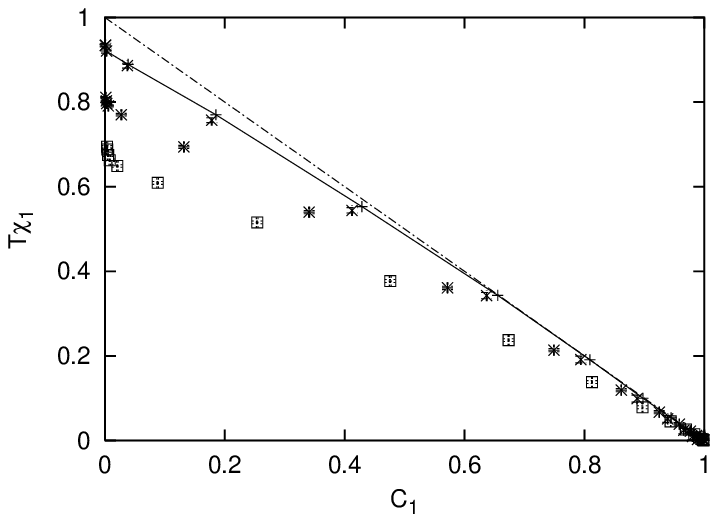,width=0.475\linewidth}&
\hspace{-0.5cm}\epsfig{figure=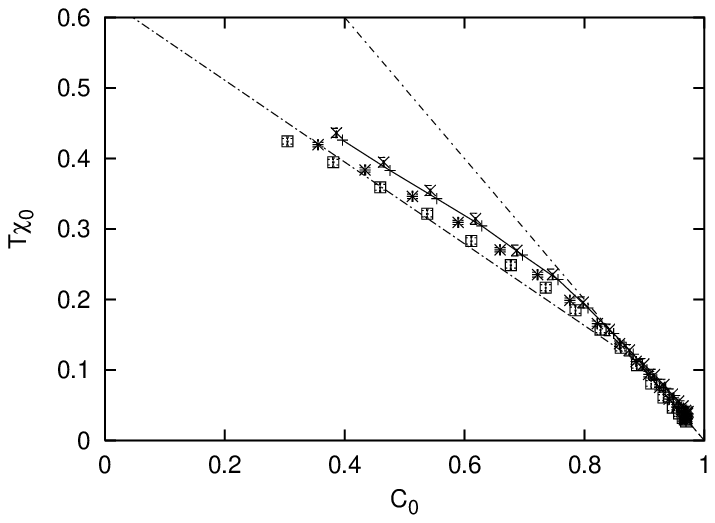,width=0.475\linewidth}
\end{tabular}
\caption{The fluctuation-dissipation plot for a type-I (left frame: site
$i=1$ on $\ha$) and for a type-II site (right frame: site $i=0$ on $\ha$)
at $T=0.4$. 
Different symbols refer to different values of the perturbing
field: $h_0 = 0.05$ ($\times$), $0.10$ ($\ast$), $0.15$ ($\Box$), and
$h_0\to 0$ extrapolation (continuous line). The dot-dashed 
lines refer to the FDT and (in the right frame) the predicted OFDR.}
\label{HFDT}
\end{figure}
In Fig. \ref{HFDT} we present the FD plots for two sites of sample $\ha$:
$i=1$ (type-I) and $i=0$ (type-II), and the three values of $h_0$
used in simulations. We expect the correlation and response 
functions to approach the zero-field limit as follows:
\begin{eqnarray}
C_i(t,t_w |h_0) =  C_i(t,t_w) + D_i(t,t_w)\, h_0^2 +O(h_0^4)\, ,
\label{Extrapolation1}\\
\chi_i(t,t_w|h_0) =  \chi_i(t,t_w) + \delta_i(t,t_w)\, h_0^2 +O(h_0^4)\, .
\label{Extrapolation2}
\end{eqnarray}
We fitted our numerical data using the above expressions (and neglecting 
next-to-leading corrections). The results of this extrapolation are 
reported in Fig. \ref{HFDT} together with the finite-$h_0$ data.

As can be verified on Fig. \ref{HFDT}, for type-II spins the $h_0\to 0$
limit is well approximated (within $2\div 3\%$) by the $h_0=0.05$,
or (within $7\div 8\%$) even by the $h_0=0.1$ result.
In what follows, we shall generally neglect finite-$h_0$ artifacts
for these sites.
On the other hand, an extrapolation of the type
(\ref{Extrapolation1})-(\ref{Extrapolation2}) is necessary for type-I spins. 
%
%
\subsubsection{Glassy degrees of freedom}
\label{3SpinGlassySection}

In this Section we focus on type-II sites, which remain out-of-equilibrium 
on long time scales.
\begin{figure}
\epsfig{figure=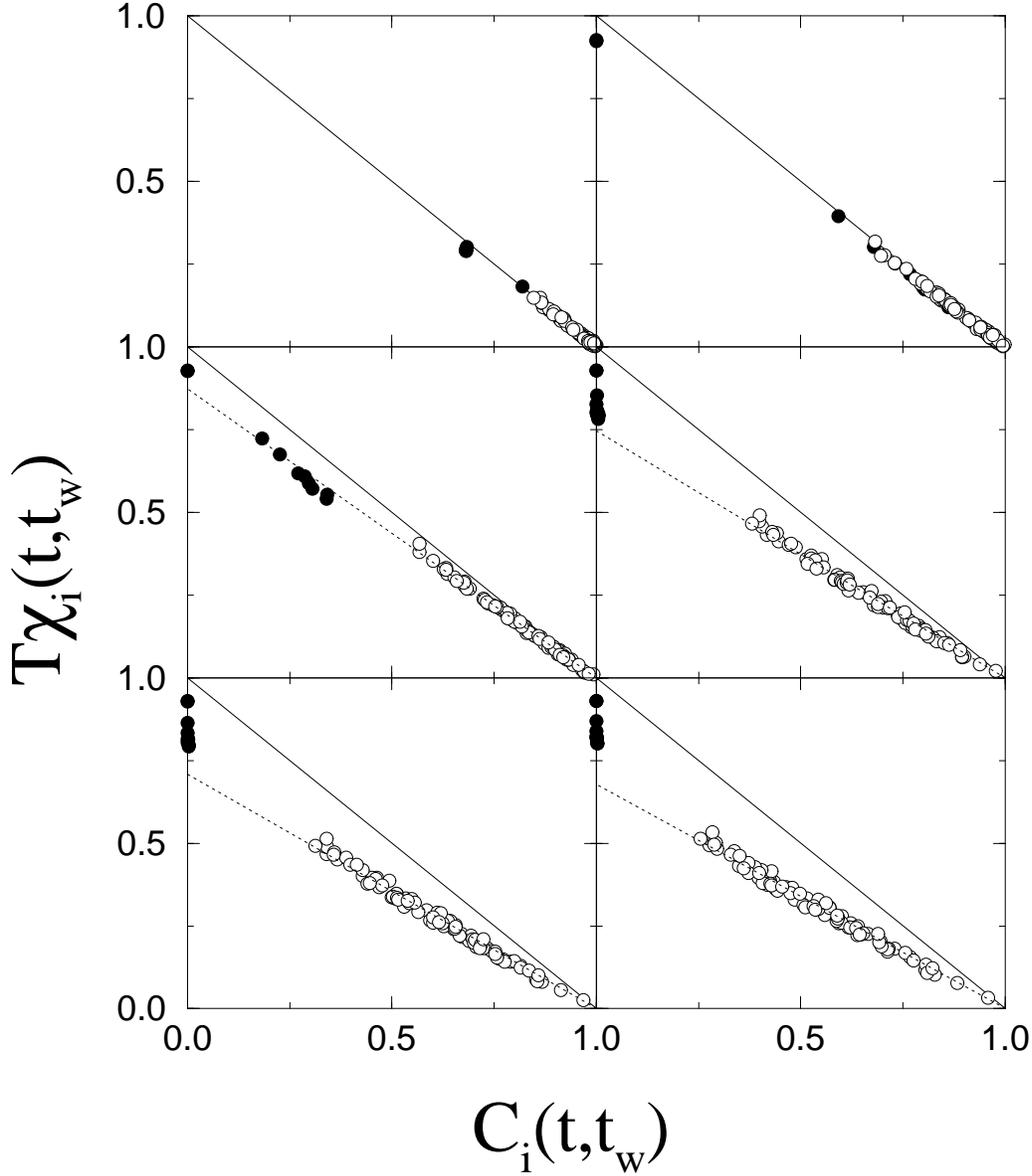,angle=-90,width=0.8\linewidth}
\caption{Movie plot for sample $\ha$ of Sec.~\ref{DiscontinuousSection}. 
Here we use $h=0.1$, $T=0.4$, and 
$t_w=10^4$. The different frames correspond to (from left to right
and top to bottom): $\Delta t=2^4$, $2^9$, $2^{12}$, $2^{15}$, $2^{16}$, and
$2^{17}$. Black and white circles refer, respectively, to type-I 
and type-II sites. Continuous lines correspond to ordinary
FDT $\chi_i = (1-C_i)/T$, while dotted ones 
are fits to a modified relation $\chi_i = (1-C_i)/T_{\rm movie}$.
We get $T_{\rm movie} = 0.459$ (for $t=2^{12}$), $0.536$ ($t=2^{15}$),
$0.564$ ($t=2^{16}$), $0.590$ ($t=2^{17}$).}
\label{Movieh01T04}
\end{figure}
In Fig. \ref{Movieh01T04} we reproduce the correlation
and response functions of {\it all} the spins of sample $\ha$ in
a movie plot.
We fix $t_w=10^4$ and watch the single-spin correlation
and response functions,  as the system evolves, i.e.
as $t$ grows. The behavior can be described as follows: 
$(i)$ for small $t$, all the points $(C_i,\chi_i)$ stay on the
fluctuation dissipation line $T\chi_i = 1-C_i$, type-I and type-II spins
cannot be distinguished; $(ii)$ as $t$ grows, type-I spins reveal to 
be ``faster'' than type-II ones and move rapidly toward the $C=0$,$\chi=1$
corner; $(iii)$ just after this, type-II spins move out of the FDT
relation, all together\footnote{Here expressions like
``all together'' must be understood as ``on the same time scale in
the aging limit''.
Let us, for instance, assume that two spins $i$ and $j$ begin 
to violate FDT after times $t_i(t_w)$ and $t_j(t_w)$. We say that
they  begin to violate FDT ``together'' if  
$\lim_{t_w\to\infty}h(t_i)/h(t_j)=1$ ($h(t)$ being the appropriate
time parametrization function).}; $(iv)$ type-II keep evolving 
in the $C$-$\chi$ plane but, amazingly, they stay, at each time on a
unique (moving) line passing through $C=1$, $\chi=0$. 

On the same graphs, in  Fig. \ref{Movieh01T04}, we show
the results of a fit of the type
\begin{eqnarray}
\chi_i(t,t_w) = \frac{1}{T_{\rm movie}(t,t_w)}[1-C_i(t,t_w)]\, .
\label{SGFittingTemperature}
\end{eqnarray}
The fit works quite well: it allows to define a new effective temperature:
the ``movie'' temperature $T_{\rm movie}(t,t_w)$. 
The thermometrical interpretation of $T_{\rm movie}(t,t_w)$ 
will be discussed in Sec. \ref{ThermoApp}. 
$T_{\rm movie}(t,t_w)$ increases with $t$ at 
fixed waiting time $t_w$. Notice the difference between 
this formula, and Eq. (\ref{FittingTemperature}) which we argued to 
hold for coarsening systems. The organization of heterogeneous degrees of
freedom in the $\chi$-$C$ plane is strongly dependent upon the nature
of the physical system as a whole.

The cautious reader will notice a few discrepancies between the above
description and the data in Fig. \ref{Movieh01T04}.
Type-I spins reach the $(C=0,\chi=1)$ corner slightly after
type-II ones move out of the FDT line. A careful check shows that this 
is a finite-$t_w$ artifact. Moreover, for large times, they stay slightly 
below the FDT line. This phenomenon has been discussed in Sec. 
\ref{FieldSection} and proved to be a finite-$h_0$ effect.

\begin{figure}
\begin{tabular}{cc}
\epsfig{figure=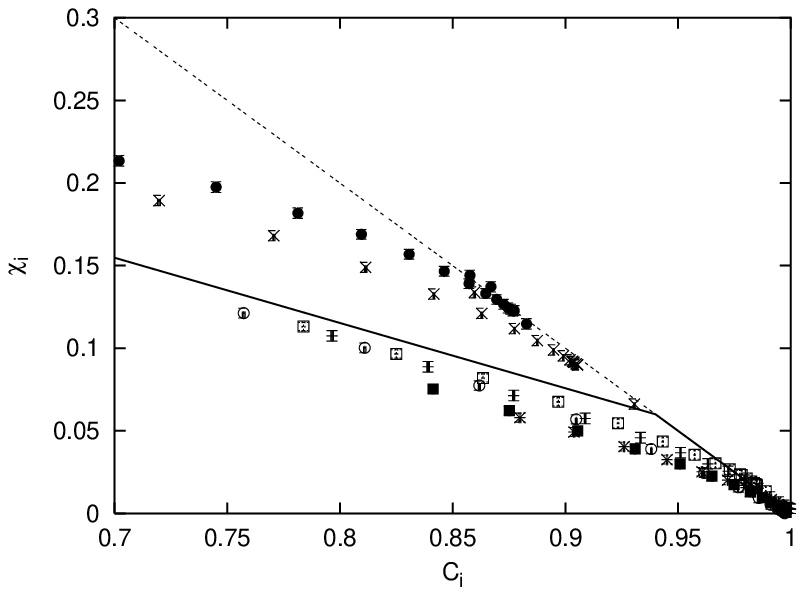,width=0.45\linewidth} &
\epsfig{figure=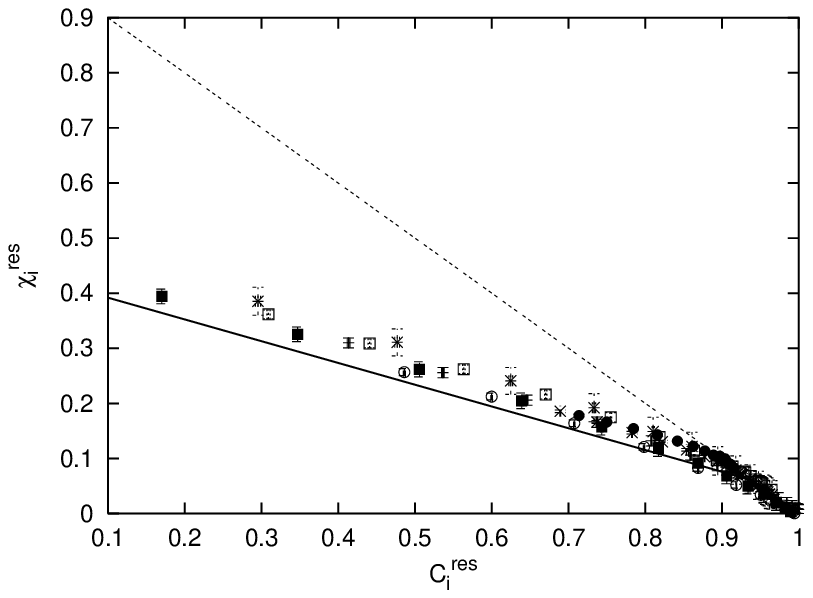,width=0.45\linewidth}
\end{tabular}
\caption{On the left, the FD plot for seven different sites:
$i=10$ ($+$), $11$ ($\circ$), $12$ (filled squares), $13$ ($\Box$), 
$16$ ($\times$), $17$ ($\ast$), $18$ ($\bullet$). 
On the right, a collapse plot of the same data,
cf. Eq. (\ref{RescaledGlass}). Here $T=0.3$, $h_0=0.1$, $t_w=10^4$,
and we use $\overline{q} = 0.94$.}
\label{Rescaling03}
\end{figure}

\begin{figure}
\begin{tabular}{cc}
\begin{tabular}{c}
\vspace{-6.cm}\\
\epsfig{figure=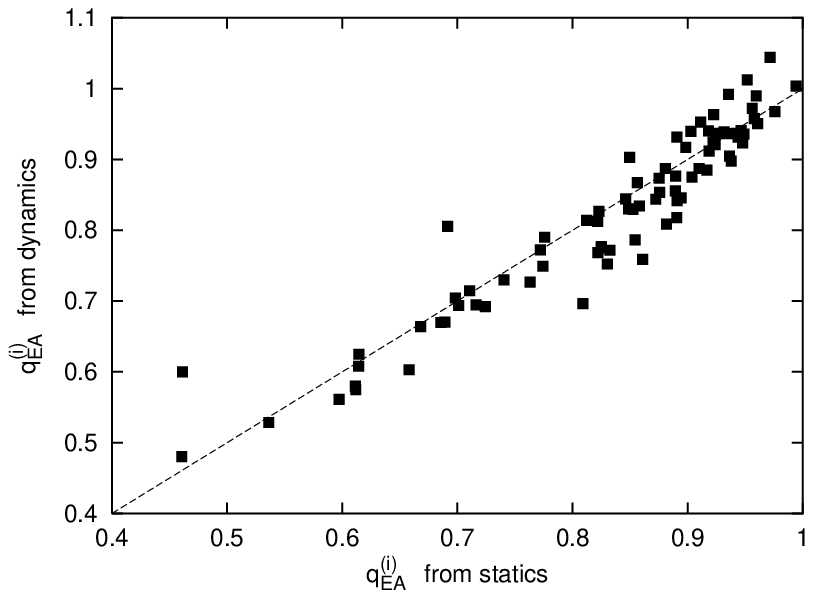,width=0.45\linewidth}
\end{tabular}
&\epsfig{figure=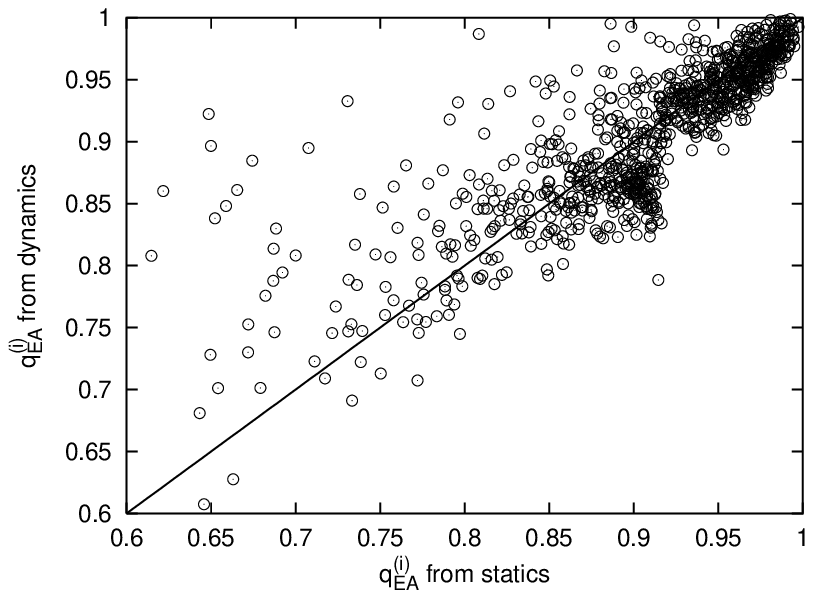,width=0.45\linewidth}
\end{tabular}
\caption{Correlation between the theoretical prediction for the local
EA parameters and the results of out-of-equilibrium simulations.
On the left we show the data for sample $\ha$ 
($T=0.5$, $t_w=10^3$, and $h_0=0.1$), on the right for sample $\hb$
($T=0.4$, $t_w=10^3$, and $h_0=0.1$).}
\label{LocalEA_3spin}
\end{figure}
\begin{figure}
\begin{tabular}{cc}
\epsfig{figure=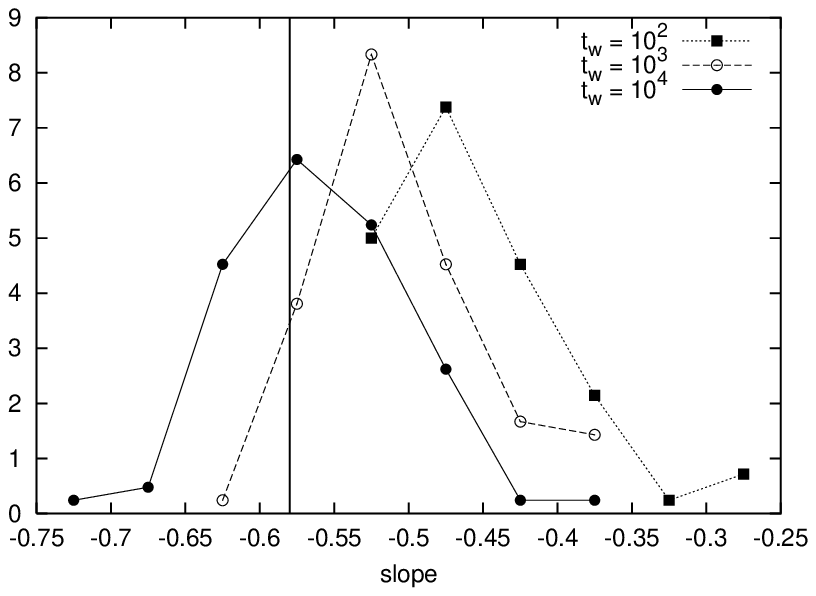,width=0.45\linewidth}&
\epsfig{figure=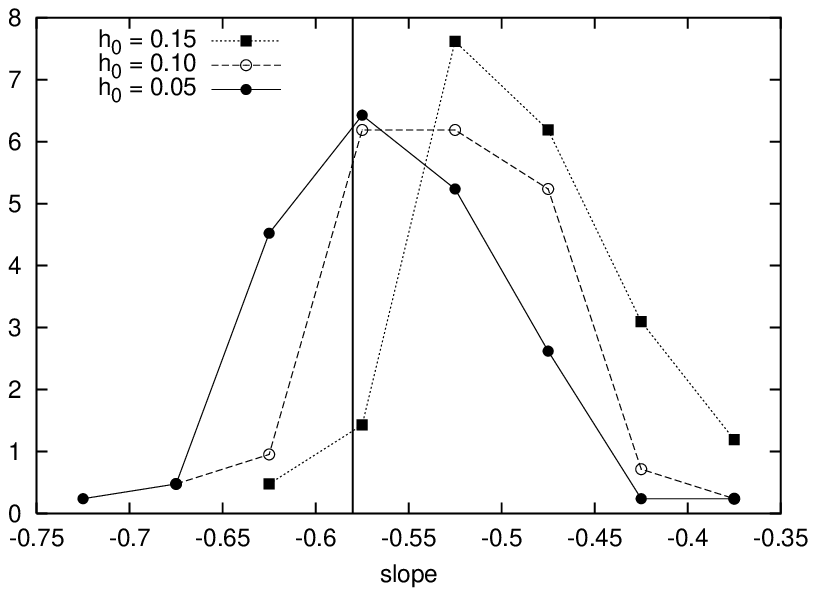,width=0.45\linewidth}
\end{tabular}
\caption{Distribution of the slopes of single-site OFDR's for $T=0.4$. 
The vertical lines correspond to the theoretical prediction for the
1RSB parameter $m_{\rm th}$. On the left (right) we kept 
$h_0=0.05$ ($t_w=10^4$) fixed.}
\label{Slopes_3spin}
\end{figure}
Let us now consider the local OFDR's, and compare the dynamical results 
with the static 1RSB prediction (\ref{OneStepChi}).
A preliminary check was given in Fig. \ref{Site0_example}.
In Fig. \ref{Rescaling03}, left frame,
we reproduce the $\chi_i$ versus $C_i$ curves for 7 type-II sites.
They are superimposed for short times (quasi-equilibrium regime) 
and spread at later times (aging regime), but remaining roughly parallel
to each other. If the static prediction (\ref{OneStepChi}) holds,
we can collapse the various $\chi_i[C_i]$ curves by properly rescaling
$\chi_i$ and $C_i$. A particular form of rescaling, which is quite 
natural for coarsening systems, was used in Sec. 
\ref{NumericalCoarseningSection}, cf. Eqs. (\ref{FerroRescaled}).
It turns out that, in this case, a better collapse can be obtained by
using the definition below:
\begin{eqnarray}
C_i^{\rm res}= 1-\frac{1-\overline{q}}{1-\qith}(1-C_i)\, ,\;\;\;\;\;\;
\chi_i^{\rm res} = \frac{1-\overline{q}}{1-\qith}\chi_i\, ,
\label{RescaledGlass}
\end{eqnarray}
where $\overline{q}$ is a reference overlap (which can be chosen freely).
In Fig. \ref{Rescaling03},
right frame, we plot $\chi_i^{\rm res}$ and $C_i^{\rm res}$ for the same
7 spins as before, computing the $\qith$ with the SP${}_T$ algorithm.
Note that there is no fitting parameter in this scaling plot.

It can be interesting to have a more general look at the 
statics-dynamics relation. In order to make a comparison, we 
fitted\footnote{While fitting the data we restricted ourselves to the
$C_i(t+t_w,t_w)<\qea^{(i)}$ time range. In practice we took
$t\ge 2^7$, $2^{10}$, and $2^{13}$, respectively for 
$t_w=10^2$, $10^3$, and $10^4$.} 
the single-site $\chi_i$-versus-$C_i$ data to the theoretical
prediction (\ref{OneStepChi}). 
The results for the two fitting parameters $\qea^{(i,{\rm fit})}$ and 
$m^{(i,{\rm fit})}$, are compared in Figs. \ref{LocalEA_3spin},
\ref{Slopes_3spin} with the outcome of the SP${}_T$ algorithm. 
Although several sources of error affect
the determination of the EA parameters 
from dynamical data, the agreement is quite satisfying.

\begin{figure}
\hspace{3.5cm}\epsfig{figure=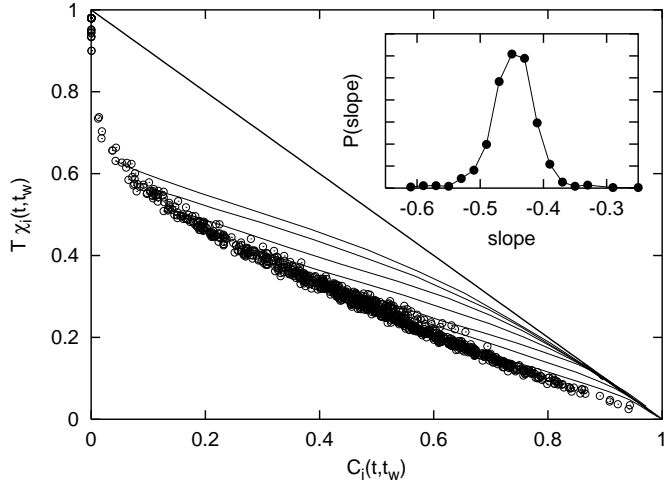,angle=0,width=0.5\linewidth}
\caption{Movie plot for sample $\hb$ ($T=0.4$, $h_0=0.1$): we show 
the position of all the degrees of freedom in the $\chi$-$C$ plane, for
$t_w = 10^3$ and $\Delta t =2^{16}$. The thin continuous lines are 
the FD plots for a few selected sites (in this 
case $\Delta t$ varies between $0$ and $2^{16}$). In the inset: 
the histogram of slopes of the FD curves in the out-of-equilibrium regime.}
\label{MovieN1000}
\end{figure}	
\begin{figure}
\centerline{\epsfig{figure=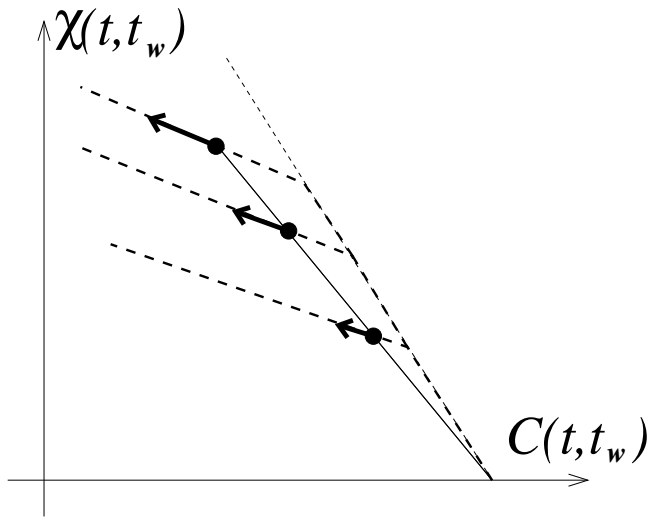,angle=0,width=0.5\linewidth}}
\caption{Qualitative picture of aging dynamics 
for discontinuous glasses. The black circles correspond 
to three different spins for a given pair of (large) times $(t,t_w)$.
Notice that, for intermediate times $\Delta t\sim t_w$ they stay on the 
same line passing through the point $(C=1,\chi=0)$.
As $t$ grows, they move with parallel velocities (arrows).
Along the time, each of them describes a different fluctuation-dissipation
curve (dashed lines).}
\label{GeneralPictureDiscontinuous}
\end{figure}
In the above paragraphs we stressed two properties of the
aging dynamics of the model (\ref{ThreeSpinHamiltonian}):
the alignment in the movie plots, cf. Fig. \ref{Movieh01T04} 
and Eq. (\ref{SGFittingTemperature}), and the OFDR (\ref{OneStepChi}).
Let us notice that these two properties 
are not compatible at all times $(t,t_w)$. In fact we expect our
model to verify the weak ergodicity-breaking condition 
$\lim_{t\to\infty}C_i(t,t_w)=0$. 
Therefore, in this limit, the alignment (\ref{SGFittingTemperature})
cannot be verified unless the $\chi_i$ become site-independent. 
On the other hand, this would invalidate the OFDR (\ref{OneStepChi}).

One plausible way-out to this contradiction 
is that Eq. (\ref{SGFittingTemperature}) breaks down
at large enough times. 
How this may happen is well illustrated by the numerical data concerning 
sample $\hb$ shown in Fig. \ref{MovieN1000}. It is quite clear that the simple
law (\ref{SGFittingTemperature}) no longer holds.
Nevertheless, it remains a very good 
approximation for the sites with a large EA parameter $\qei\gtrsim 0.5$.
Moreover, it seems that the points corresponding to different
sites still lie on the same curve in the $\chi$-$C$ plane, although this 
curve is not a straight line as in Eq. (\ref{SGFittingTemperature}). 
We shall further comment on this point in Sec.
\ref{DiscussionSection}.

The general picture which holds at intermediate 
times (or large $\qei$'s) for discontinuous glasses is
summarized in Fig. \ref{GeneralPictureDiscontinuous}. This
should be compared with Fig. \ref{FerroGeneralPicture}, which refers 
to coarsening systems.
%
%
\section{Continuous glasses}
\label{ContinuousSection}

The Viana-Bray model \cite{VianaBray} is a prototypical example of continuous 
spin glass. It is defined by the Hamiltonian
\begin{eqnarray}
H(\sigma) = -\sum_{(ij)\in {\cal G}} J_{ij}\sigma_i\sigma_j\, ,
\end{eqnarray}
where the graph ${\cal G}$ is constructed by randomly choosing 
$M$ among the $N(N-1)/2$ couples of spins, and the couplings
$J_{ij}$ are independent identically distributed random variables.
The average connectivity of the graph is given by $c=2M/(N-1)$.
If we assume that the couplings distribution is even, the phase diagram 
of this model is quite simple \cite{VianaBray,GuerraToninelli}. 
For $c<1$ the interaction graph
does not percolate and the model stays in its paramagnetic phase at all finite 
temperatures. For $c>1$ the graph percolates and the giant component
undergoes a paramagnetic-spin glass phase transition. The critical temperature 
is given by the solution of the equation $\mathbb{E}_J(\tanh \beta J)^2=1/c$.
Below the critical temperature, a finite Edwards-Anderson
parameter $\qea$ develops continuously from zero.

We considered three samples of this model: hereafter they will be denoted
as $\ga$, $\gb$ and $\gc$. The interaction graph and the signs of 
the interactions $J_{ij}$ were the same for $\ga$ and $\gb$:
in particular we used $N=1000$ and $M=1999$, i.e. $c\approx 4$,
and chosen the interaction signs to be $\pm 1$ with equal probabilities.
The two samples differ only in the strength of the couplings. 
While in $\ga$
we used $|J_{ij}|=1$, in $\gb$ we took $|J_{ij}| = kJ_0$, where
$k\in\{1,\dots,10\}$ with uniform probability distribution and
$J_0 = 0.161164$\footnote{This distribution has a three nice properties:
$(i)$ being discrete, it allows to precompute the table of 
Metropolis acceptance probabilities; $(ii)$ since 
we will use temperatures $T\gg J_0$, it is a good approximation 
to a continuum distribution; $(iii)$ it has ${\mathbb E}_J(J^2)\approx 1$.}.
We made this choice in order to check the effects of 
degenerate coupling strengths on the aging dynamics.
The sample $\gc$ was instead much larger:
we used $N=10000$, $M=20190$ (once again $c\approx 4$) 
and $J_{ij} = \pm 1$ with equal probabilities.
The critical temperatures for $c=4$ and the two coupling distributions 
used here are $T_c\approx 1.8204789$ (for $\ga$ and $\gc$) 
and $1.6717415$ ($\gb$).

The glassy phase of the VB model is thought to be characterized by
FRSB. Nevertheless we can
use the SP${}_T$ algorithm to compute a one-step approximation to the local 
overlaps and the local OFDR's. Of course, such an approximation will have the
simple two time-sectors form, see Eq. (\ref{OneStepChi}), instead of the
expected infinite-time-sectors behavior.
However the situation is not that simple because of two problems:
\begin{itemize}
\item We expect, in analogy with the Sherrington-Kirkpatrick model
\cite{CugliandoloKurchanSK},
the dynamics of this model to reach the equilibrium free-energy in the long 
time limit. It is not clear whether a better approximation
to the correct OFDR is obtained by using the threshold value
$m_{\rm th}$ or the ground-state value $m_{\rm gs}$ of the 1RSB parameter.
\item The SP${}_T$ algorithm does not converge. After a fast transient 
the probability distributions of local fields oscillate indefinitely.
This is, plausibly, a trace of FRSB.
\end{itemize}
The first problem does not cause great trouble because the two 
determinations of $m$ are, generally speaking, quite close. On the other hand, 
we elaborated two different way-out to the second one: $(i)$
To force the local field distributions to be symmetric (which can be expected 
to be true on physical grounds): this assures convergence. 
$(ii)$ To average the local EA parameters
over sufficiently many iterations of the algorithm.

While the approach $(i)$ seems physically more sound, it
underestimates grossly the $q^{(i)}_{\rm EA}$'s. The approach $(ii)$,
which will be adopted in our analysis, gives much more reasonable
results.  Notice that the authors of
Refs.\cite{MezardParisiBethe,MezardParisiBetheZeroT} followed the same
route.  In their calculation, they faced no problem of convergence.
In fact they required convergence in distribution, while we require
convergence site-by-site.
%
%
\subsection{Numerical results}

Most of our simulations were run at temperature $T=0.5$, and with
$h_0=0.1$. We used waiting times $t_w=10^2$, $10^3$, $10^4$, and,
respectively $\Delta t_{\rm MAX}=2^{14}$, $2^{16}$, $2^{18}$.
In the Table below we summarize the statistics used in each case.

\vspace{0.3cm}
\begin{tabular}{|c||c|c|c|}
\hline
  & $t_w=10^2$ & $t_w=10^3$ & $t_w=10^4$  \\
\hline
\hline
$\ga$ & $10^6$ & $4\cdot 10^5$ & $5.5\cdot 10^5$ \\
\hline
$\gb$ & $6\cdot 10^5$ & $1.5\cdot 10^6$ & $\ast$ \\
\hline
\hline
$\gc$ & $6\cdot 10^5$ & $\ast$ & $\ast$ \\
\hline
\end{tabular}
\vspace{0.3cm}

Moreover we simulated $N_{\rm stat} = 4.2\cdot 10^5$ Metropolis trajectories
at temperature $T=0.4$ on sample $\ga$ with 
$t_w=10^4$ and $\Delta t_{\rm MAX} = 2^{18}$.

\begin{figure}
\centering\epsfig{figure=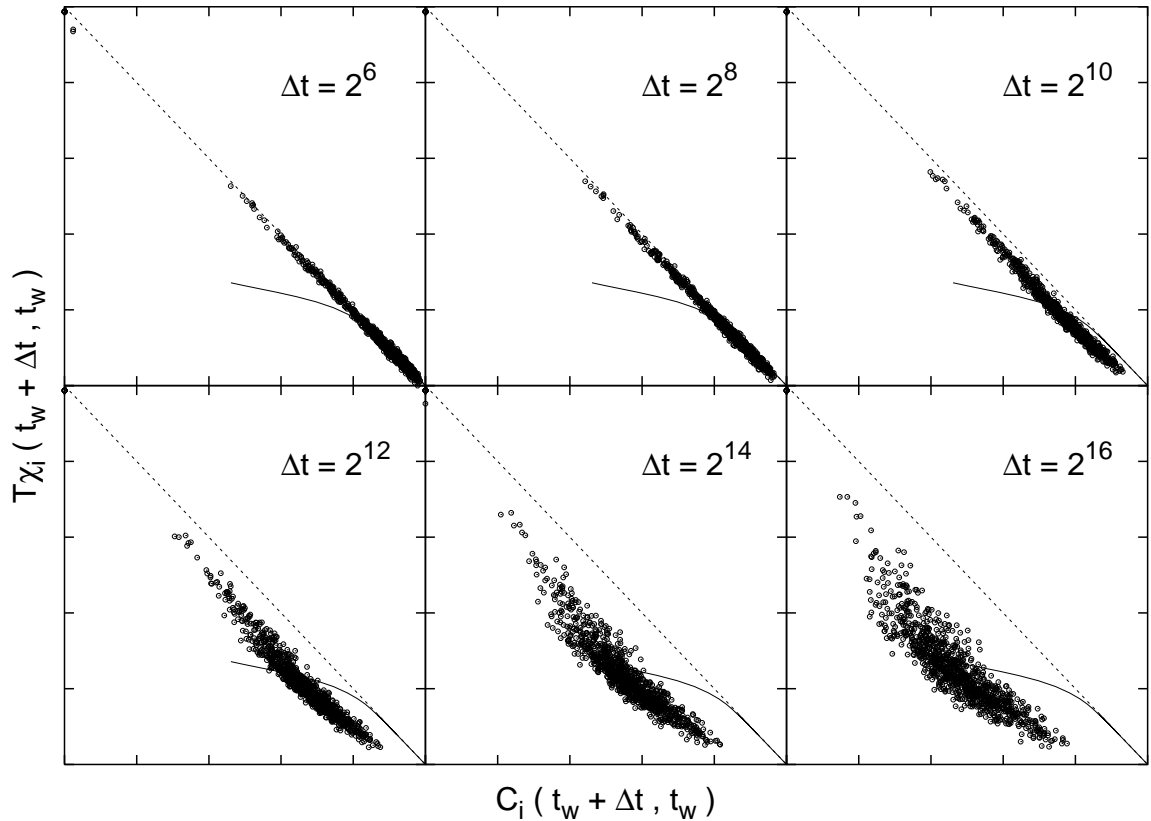,angle=-90,width=0.9\linewidth}
\caption{Single-spin correlation and response functions for the sample 
$\ga$ (VB model) for $T=0.5$, $h_0 = 0.1$, and $t_w = 10^4$. 
The continuous line and full circle refer to the global 
correlation and response.}
\label{MovieVB}
\end{figure}	
\begin{figure}
\centering\epsfig{figure=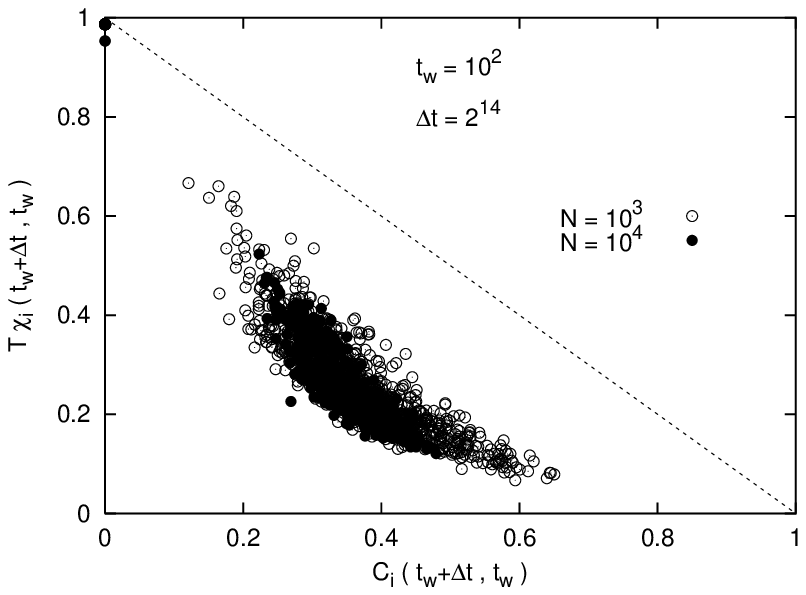,angle=0,width=0.75\linewidth}
\caption{Single-spin correlation and response functions: here
we compare the results obtained on samples $\ga$ ($\circ$) and 
$\gc$ ($\bullet$) which are of different sizes.}
\label{CloudsFig}
\end{figure}	
In Fig. \ref{MovieVB} we show the movie plot of sample 
$\ga$ for $t_w=10^4$. As in the previous Sections, the local 
correlation and response functions are strongly heterogeneous:
the global two-times functions give just a rough idea of the dynamics of
the system.
Moreover all the points quit the FDT line on the same time scale
in the aging limit (cf. Sec. \ref{3SpinGlassySection}). 
However, their behavior in the aging regime does not fit any of
the alignment patterns we singled-out in the case of coarsening systems,
cf. Eq. (\ref{FittingTemperature}) and Fig. \ref{FerroGeneralPicture}, 
or discontinuous glasses, cf. Eq. (\ref{SGFittingTemperature}) 
and Fig. \ref{GeneralPictureDiscontinuous}.
We repeated the same type of analysis for the numerical data
obtained on sample $\gc$. 
In this case, see Fig. \ref{CloudsFig}, the  points corresponding to local 
correlation and response functions are much less spread in the
$\chi$-$C$ plane. Therefore our simulations are quite inconclusive 
on the possibility of defining a ``movie'' temperature as
in Eq. (\ref{SGFittingTemperature}). To settle the question, simulations 
on larger samples are probably necessary.
\begin{figure}
\begin{tabular}{cc}
\hspace{-1.cm}
\epsfig{figure=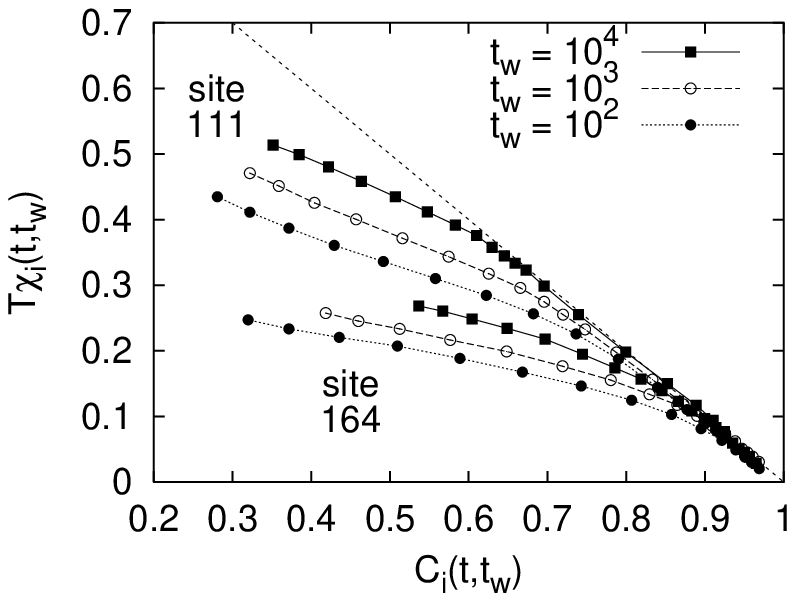,angle=0,width=0.5\linewidth}&
\epsfig{figure=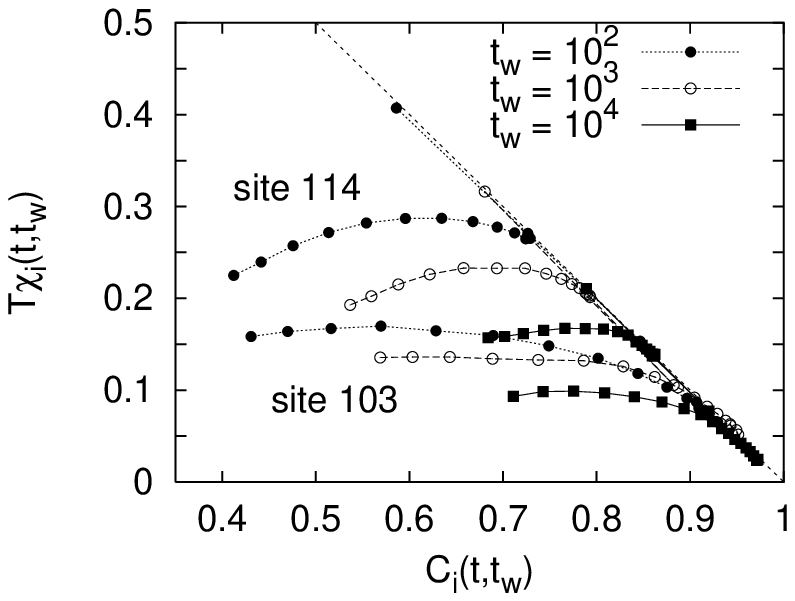,angle=0,width=0.5\linewidth}
\end{tabular}
\caption{FD plot for a few selected sites of sample $\ga$ ($T=0.5$,
$h_0 = 0.1$). Notice the completely different behaviors of the sites
in the two frames. The sites in the left frame, with connectivity $1$
(site $111$) and $3$ (site $164$), look like a ``glassy'' system. The
ones on the right, with connectivity $4$ (site $103$) and $2$ (site
$114$), look like a ``coarsening'' system.}
\label{FDVB}
\end{figure}	

Numerical results on sample $\ga$ are also deceiving for what concerns
local OFDR's, cf. Fig. \ref{FDVB}. 
It seems that the local FD plots
depend strongly upon the waiting time and the particular site. Moreover
the slopes of this plots (for a given couple $(t_w,\Delta t)$) change 
from site to site.
\begin{figure}
\centerline{\epsfig{figure=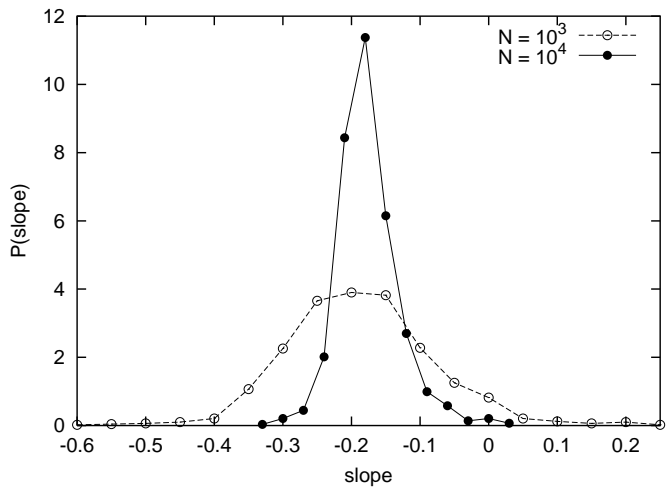,angle=0,width=0.5\linewidth}}
\caption{Distributions of slopes of the local FD
plots in the aging regime. The two curves refer to sample $\ga$ ($\circ$)
and $\gc$ ($\bullet$), which are of different sizes.}
\label{HistPendVB}
\end{figure}	
These effects are much smaller in sample $\gc$. In Fig.
\ref{HistPendVB} we consider the distribution of slopes of local
FD plots for samples $\ga$ and $\gc$. We computed 
the slopes by fitting the aging part of the plot to the one-step
form (\ref{OneStepChi}).
\begin{figure}
\centerline{\epsfig{figure=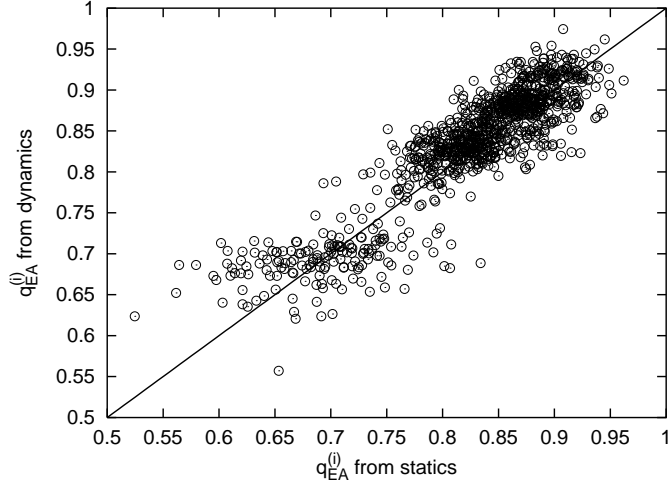,angle=0,width=0.5\linewidth}}
\caption{Local EA parameters on sample $\gc$. Numerical results,
obtained by fitting the aging part of the FD plots,
are compared with the outcome of the SP${}_T$ algorithm.}
\label{LocalQeaVB}
\end{figure}	

By the same fitting procedure we extracted the local EA parameters.
The comparison with the predictions of the SP${}_T$ algorithm, cf.
Fig. \ref{LocalQeaVB} is quite satisfying. Notice that, both in analyzing the
numerical data and in using the SP${}_T$ algorithm, we are adopting
a 1RSB approximation, cf. Eq. (\ref{OneStepChi}), to the real OFDR.
The slopes considered in Fig. \ref{HistPendVB} should therefore
be understood as {\it average} slopes in the aging regime.
We expect the systematic error induced by this approximation to be small.

The arguments of Ref. \cite{ParisiLocal} imply that the  slopes
(effective temperatures) of the OFDR's for different degrees of freedom 
should be identical. This conclusion is valid only 
in the aging window $1\ll t_w, \Delta t\ll t_{\rm erg}(N)$.
Our numerical data, cf. Fig. \ref{HistPendVB}, suggest a clear trend
confirming this expectation. Nevertheless, they show large finite-size 
effects due, arguably, to a mild divergence of $t_{\rm erg}(N)$ with $N$:
the smaller ($N=10^3$) samples begin to equilibrate during the simulations.
This is quite different from what happens with discontinuous
glasses, cf. Sec. \ref{DiscontinuousSection}. In that case, we did not
detect any evidence of equilibration even in sample $\ha$ ($N=10^2$).
A better understanding of the scaling of  $t_{\rm erg}(N)$ in different 
classes of models would be welcome.
%
%
\section{Weakly interacting spins}
\label{WeaklySection}

We lack of analytical tools for studying the dynamics of 
diluted mean-field spin glasses (for some recent work see
Refs. \cite{SemerjianCugliandolo,JonaBirthday,WalkSatMonasson,WalkSatWeigt}).
This makes  somehow ambiguous the interpretation of many numerical results.
For instance, the identity of effective temperatures for different spins,
although consistent with our data, see Figs. \ref{Slopes_3spin} and 
\ref{HistPendVB}, could still be questioned. 
This would contradict the general 
arguments of Refs. \cite{ParisiLocal,CugliandoloKurchanPeliti}.
Even more puzzling is the definition of a ``movie'' temperature 
along the lines of Eq. (\ref{SGFittingTemperature}).
Such a definition seems to be consistent only in some particular models and
time-regimes.
In this Section we want to point out a simple perturbative calculation 
which supports the identity of single-spin effective temperatures, in 
agreement with the standard wisdom. Moreover it give some intuition on
the range of validity of the definition (\ref{SGFittingTemperature}).

Let us consider a generic diluted mean-field 
spin glass with $k$-spin interactions:
\begin{eqnarray}
H(\sigma) = -\sum_{\ua \in {\cal H}} J_{\ua} 
\sigma_{\alpha_1}\cdot\dots\cdot\sigma_{\alpha_k}\, .
\end{eqnarray}
Here $\ua = \{\alpha_1,\dots,\alpha_k\}$ 
is a $k$-uple of interacting spins, and ${\cal H}$ is
a $k$-hypergraph, i.e. a set of $M$ such $k$-uples. 

Let us focus on a particular site, for instance $i=0$, 
and  assume that it is weakly coupled to its 
neighbors. It is quite natural to think that its response and correlation
functions can be related to the response and correlation functions
of the neighbors. To the lowest order this relation reads:
\begin{eqnarray}
C_0^{\rm ag}(t,t_w) & = & \sum_{\ua \ni 0} (\tanh \beta J_{\ua})^2
\prod_{i\in\ua \backslash 0} C_i^{\rm ag}(t,t_w) + O(\beta^4J^4)\, ,
\label{PertGeneral_Corr}\\
R_0^{\rm ag}(t,t_w) & = & \sum_{\ua \ni 0} (\tanh \beta J_{\ua})^2
\sum_{i\in\ua \backslash 0} R_i^{\rm ag}(t,t_w)
\prod_{j\in\ua \backslash \{0,i\}} C_j^{\rm ag}(t,t_w)+ O(\beta^4J^4)\, .
\label{PertGeneral_Resp}
\end{eqnarray}
We shall not give here the details of the derivation.
The basic idea is to use an appropriate  dynamic 
generalization of the cavity method \cite{SpinGlass,DynamicCavity}. 
As for static calculations \cite{MarcRiccardo}, 
this approach gives access to single-site quantities for a given 
disorder realization.
Notice that Eq. (\ref{PertGeneral_Corr}) 
can be easily obtained by assuming that the spin $\sigma_0$ does not
react on its neighbors. This is not the case for Eq. (\ref{PertGeneral_Resp}). 

Equation (\ref{PertGeneral_Corr})
implies a relation between local Edwards-Anderson 
parameters: 
\begin{eqnarray}
q_{\rm\scriptscriptstyle EA}^{(0)}
& = & \sum_{\ua \ni 0} (\tanh \beta J_{\ua})^2
\prod_{i\in\ua \backslash 0} \qei + O(\beta^4J^4)\, .
\end{eqnarray}
In the $k=2$ (Viana-Bray) case, we can derive
from Eq. (\ref{PertGeneral_Resp}) a simple relation between 
the integrated responses:
\begin{eqnarray}
1-T\chi_0^{\rm ag}(t,t_w) & = & \sum_{i\in \partial 0}(\tanh \beta J_{0i})^2 
[1-T\chi_i^{\rm ag}(t,t_w)]+ O(\beta^4J^4)\, ,\label{PertVianaBray2}
\end{eqnarray}
where $\partial i_0$ denote the set of neighbors of the spin $i_0$. 
In the general ($k>2$) case Eq. (\ref{PertGeneral_Resp}) cannot be integrated 
without further assumptions.

\begin{figure}
\begin{tabular}{cc}
\epsfig{figure=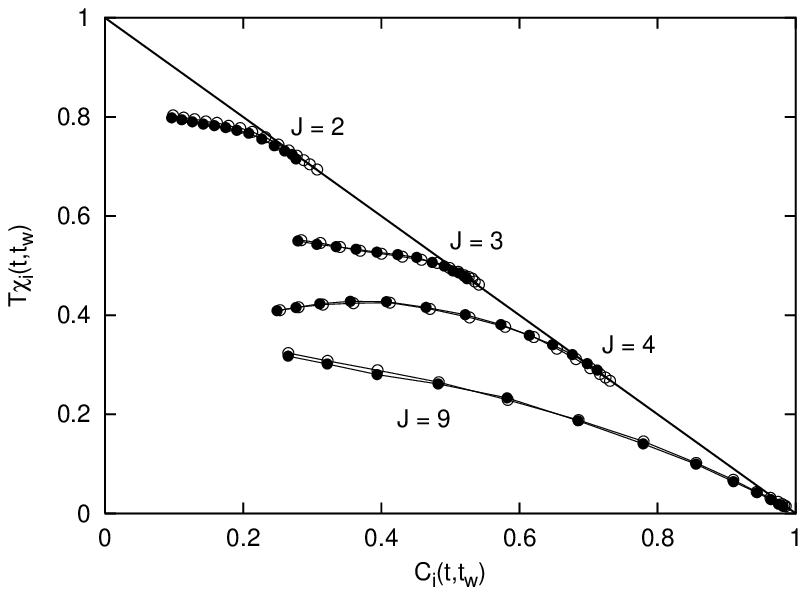,angle=0,width=0.45\linewidth}&
\epsfig{figure=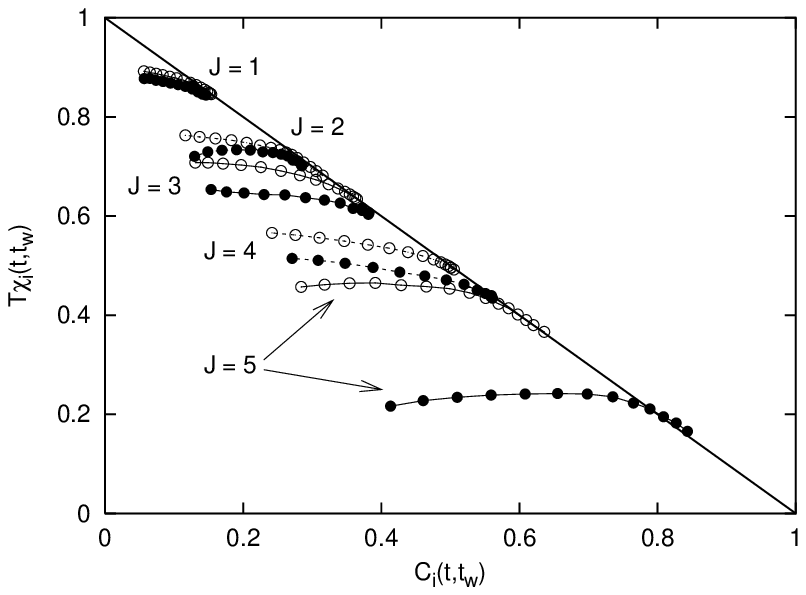,angle=0,width=0.45\linewidth}
\end{tabular}
\caption{FD plots for a few weakly-interacting
spins: numerical results ($\bullet$) and outcomes of the perturbative
formulae (\ref{PertGeneral_Corr}) and (\ref{PertVianaBray2}) ($\circ$).
We consider connectivity-1 sites on the left, and
connectivity-2 sites on the right.}
\label{WeaklyFig}
\end{figure}	

We checked the above relations on our numerical data for the Viana-Bray
model. Sample $\gb$ is particularly suited for this task,
since we can choose spins whose interactions have a varying strength.
In Fig. \ref{WeaklyFig}, we consider a few spins with connectivity one
and two, and compare their correlation and response functions with the outcome
of Eqs. (\ref{PertGeneral_Corr}) and (\ref{PertVianaBray2}). 
Of course, the perturbative formulae are well verified only for small
couplings.
For connectivity-2 sites we have plotted in Fig.~\ref{WeaklyFig} only
those with coupling of the same strenght, since spins with 2 couplings
of very different strengths behave very similarly to connectivity-1 spins.

Let us now discuss some implications of Eqs. (\ref{PertGeneral_Corr}),
(\ref{PertGeneral_Resp}).
If we define the fluctuation-dissipation ratio as 
$X_i(t,t_w)\equiv T R_i^{\rm ag}(t,t_w)/\partial_{t_w}C_i^{\rm ag}(t,t_w)$,
we get:
\begin{eqnarray}
X_0(t,t_w) \approx \frac{\sum_{\ua\ni 0}\sum_{i\in\ua\backslash 0}
W_{\ua,i }(t,t_w)
X_i(t,t_w)}{\sum_{\ua\ni 0}\sum_{i\in\ua\backslash 0}
W_{\ua,i }(t,t_w)}\, , \label{WeightedAverage}
\end{eqnarray} 
where 
\begin{eqnarray}
W_{\ua,i }(t,t_w) \equiv (\tanh\beta J_{\ua})^2\partial_{t_w}C_i(t,t_w)
\prod_{j\in\ua \backslash \{i,0\}} C_j(t,t_w)
\end{eqnarray} 
are positive weights. Therefore, at the lowest order in perturbation
theory, the effective temperature of the spin $\sigma_0$ is a weighted 
average of the effective temperatures of its neighbors.
Let us suppose that this conclusion
remains {\it qualitatively} true beyond perturbation theory.
It follows that $X_i(t,t_w) = X(t,t_w)$ is independent of the site $i$.
In fact, if the $X_i(t,t_w)$ were site-dependent we could just 
consider a site $i_*$ such that 
$X_{i_*}(t,t_w)$ is a relative maximum and show that
Eq. (\ref{WeightedAverage}) cannot hold on such a site.
With a suggestive rephrasing we may say that 
effective temperatures must diffuse until they becomes site-independent.

Moreover, Eqs. (\ref{PertGeneral_Corr}) and (\ref{PertGeneral_Resp}) 
can be used to construct examples of weakly interacting spins which violate
the alignment in the $\chi$-$C$ plane which we encountered for 
discontinuous glasses, cf. Eq. (\ref{SGFittingTemperature})
and Fig. \ref{Movieh01T04}. 
The simplest of such examples is obtained by considering
the Viana-Bray ($k=2$) case, and assuming that the site 0 has just one 
neighbor. In this case it is immediate to show that
\begin{eqnarray}
\frac{C_0^{\rm ag}(t,t_w)}{1-T\chi_0^{\rm}(t,t_w)}\approx
\frac{C_i^{\rm ag}(t,t_w)}{1-T\chi_i^{\rm}(t,t_w)}\, ,
\end{eqnarray}
i.e. weakly interacting spins have the tendency to align as in coarsening 
systems. The reader can easily construct analogous examples for $k>2$ models. 
This suggests that the ``movie'' temperature 
(\ref{SGFittingTemperature}) is well defined uniquely for strongly
interacting and glassy systems, or, in other words, for slow-evolving
sites with a $\qei$ close to one.
%
%
\section{Discussion}
\label{DiscussionSection}

In the last two Sections we shall discuss the general properties of
single-spin correlation and response functions which emerge from the numerics.
We will give an overview of such properties in the present Section,
and precise the thermometric interpretation of some of them in the next one.

We shall focus on two-times correlation and response functions
$C_i(t,t_w)$ and $R_i(t,t_w)$ (see \cite{JonaBirthday} for a preliminary
discussion of multi-time functions) and distinguish two types of facts:
$(i)$ their scaling behavior in the large time limit; $(ii)$
the fluctuation-dissipation relations which connect correlation
and response.

\subsection{Time scaling}

Following Refs. 
\cite{CugliandoloKurchanSK,CugliandoloKurchanWeak}, we assume monotonicity
of the two-times functions:
$\partial_t C_i(t,t_w)$, $\partial_t R_i(t,t_w)\le 0$, and
$\partial_{t_w} C_i(t,t_w)$, $\partial_{t_w} R_i(t,t_w)\ge 0$. Moreover we
consider a weak-ergodicity breaking situation:
$C_i(t,t_w),\, R_i(t,t_w)\to 0$ as $t\to\infty$ for any fixed $t_w$. All these
properties are well realized within our models.

It is quite natural to assume\footnote{A similar 
statement appears in Ref.
\cite{CugliandoloKurchanLeDoussal}. Notice however two differences:
the cited authors consider correlation functions for different {\it modes}
in Fourier space, while we consider different {\it sites}; more important,
they take the average over quenched disorder, while we consider a fixed 
realization of the disorder (our statement would be trivial if we took the
disorder average).} that, 
for pair of sites $i$ and $j$ there exist 
two continuous functions $f_{ij}$ and $f_{ji}$ such that
\begin{eqnarray}
C_i(t,t_w) = f_{ij}[C_j(t,t_w)]\,,\;\;\;\;
C_j(t,t_w) = f_{ji}[C_i(t,t_w)]\, , \label{Equivalence}
\end{eqnarray}
in the $t,t_w\to \infty$ limit. Notice that we can always write
\begin{eqnarray}
C_i(t,t_w) = f_{ij}[C_j(t,t_w),t]\, .
\end{eqnarray}
We are therefore assuming that the functions $f_{ij}[C,t]$ admit a 
limit as $t\to\infty$ and that the limit is continuous.
Since $f_{ij}[C,t]$ is smooth and $\partial_{C}f_{ij}[C,t]\ge 0$,
if the limit exists it must be a continuous, non-decreasing function of
$C$. Since Eq. (\ref{Equivalence}) implies that both $f_{ij}$ and $f_{ji}$
are invertible (indeed $f_{ij}\circ f_{ji}=1$, see below) 
they must be strictly increasing.

Without any further specification,
the property (\ref{Equivalence}) is trivially {\it false}.  
Consider the example of type-I (paramagnetic)
spins in the 3-spin model studied in Sec. \ref{DiscontinuousSection}. 
If $i$ is type-I, and $j$ is type-II,  $C_i(t,t_w)\to 0$ in the 
aging regime, while $C_j(t,t_w)$ remain non-trivial: $f_{ij}[\cdot]$
cannot be inverted.
Another example, would be that of a Viana-Bray model, cf.
Sec. \ref{ContinuousSection}, such that the interactions graph has
two disconnected components.

However, both these counter-examples are somehow ``pathological''.
We can precise this intuition by noticing that Eq.
(\ref{Equivalence}) defines an equivalence relation (in mathematical sense)
between the sites $i$ and $j$. Therefore the physical system 
breaks up into {\it dynamically connected components} which are the
equivalence classes of this relation. Type-I and type-II spins in 
the 3-spin model of  Sec. \ref{DiscontinuousSection} are two
examples of dynamically connected components.  
Hereafter we shall restrict our attention to a single
dynamically connected component.
Physically, structural rearrangements occurs coherently within 
such a component.

Clearly the transition functions $\{ f_{ij}\}$ have the following
two properties:
$(i)$ $f_{ji}=f^{-1}_{ij}$, and $(ii)$ $f_{ij}=f_{ik}\circ f_{kj}$.
This implies that they can be written in the form 
$f_{ij} =f_i^{-1}\circ f_j$ 
(the proof consists in taking a reference spin $k=0$ and
writing  $f_{ij}=f_{i0}\circ f_{0j}=f_{0i}^{-1}\circ f_{0j}$). Of course
the functions $f_i$ are not unique: in particular they can be modified by
a global reparametrization $f_i\to g\circ f_i$.

Although very simple, the hypothesis (\ref{Equivalence}) has some important 
consequences. Suppose that $C_j(t,t_w)$ has $p$ discrete correlation scales
(in the sense of Refs. \cite{CugliandoloKurchanSK,CugliandoloKurchanWeak}),
characterized  by $q^{(j)}_{\alpha+1}<C_j(t,t_w)\le q^{(j)}_{\alpha}$,
for $\alpha=1,\dots,p$. Within a scale we have
\begin{eqnarray}
C_j(t,t_w)\approx 
{\cal C}^{(\alpha)}_j(h^{(j)}_{\alpha}(t)/h^{(j)}_{\alpha}(t_w))\, .
\end{eqnarray}
where $h^{(j)}_{\alpha}(t)$ is a monotonously increasing {\it time-scaling 
function}. Two times $t$ and $t_w$ belong to the same time sector if
$1<h^{(j)}_{\alpha}(t)/h^{(j)}_{\alpha}(t_w)<\infty$.

Applying the transition function $f_{ij}$ to the above equation, one 
can prove that, for each scale $\alpha$ of the site $j$,
there exists a correlation scale for the site $i$, with 
$q^{(i)}_{\alpha+1}<C_i(t,t_w)\le q^{(i)}_{\alpha}$ and $q^{(i)}_{\cdot}=
f_{ij}[q^{(j)}_{\cdot}]$. Moreover $h^{(i)}_{\alpha}(t)=h^{(j)}_{\alpha}(t)
\equiv h_{\alpha}(t)$ (up to an irrelevant multiplicative constant) and
\begin{eqnarray}
{\cal C}^{(\alpha)}_i = f_{ij}\circ {\cal C}^{(\alpha)}_j\, .
\end{eqnarray}
In summary there is a one-to-one correspondence between the correlation
scales of any two  sites. 
Notice that this is a necessary hypothesis if we want the
connection between statics and dynamics \cite{StatDyn1,StatDyn2} 
to be satisfied both at the level of global and local 
(single-spin) observables.
A spectacular demonstration of the correspondence of correlation scales 
on different sites is given by our movie plots, cf. Figs.
\ref{Ferro_FDMovie}, \ref{Movieh01T04} and \ref{MovieVB}. 
In particular such correspondence implies 
that all the $(\chi_i,C_i)$ points leave the FDT line at once.

Equation (\ref{Equivalence}) can be rephrased by saying that the
behavior of one spin ``determines'' the behavior of the whole 
system. This is compatible with the locality of the underlying dynamics
because: $(i)$ ``determines'' has to be understood in average sense;
$(ii)$ the relation (\ref{Equivalence}) is not true but in the aging limit.

\subsection{Fluctuation-dissipation relations}

On general grounds, we expect single-spin 
quantities to to satisfy site-dependent OFDR's of the type (\ref{OFDR}).
In integrated form we obtain, for large times $t,t_w$, the relation
$\chi_i(t,t_w)= \chi_i[C_i(t,t_w)]$. We think that we accumulated
convincing numerical evidence in this direction as far
as the models of Secs. \ref{CoarseningSection} (coarsening) and 
\ref{DiscontinuousSection} 
(discontinuous spin glass) are considered. The situation is more
ambiguous (and probably very hard to settle numerically) 
for the Viana-Bray model of Sec. \ref{ContinuousSection}.

Fluctuation-dissipation relations on different sites are not unrelated:
we expect \cite{ParisiLocal} 
to be able to define a site-independent effective temperature 
as follows
\begin{eqnarray}
\chi_i'[C_i(t,t_w)] = \chi_j'[C_j(t,t_w)] 
\equiv -\frac{1}{T_{\rm eff}(t,t_w)}\, .
\label{SlopesIdentity}
\end{eqnarray}
In terms of transition functions, we get 
$\chi_i'[C_i] = \chi_j'[C_j]$ when $C_i=f_{ij}[C_j]$.
As before, the numerics support this identity both for coarsening systems,
cf. Sec. \ref{CoarseningSection}, and discontinuous
glasses, cf. Sec. \ref{DiscontinuousSection}. For continuous glasses,
cf. Sec. \ref{ContinuousSection}, the situation is less definite.
In Sec. \ref{WeaklySection} we presented a perturbative calculation which
support Eq. (\ref{SlopesIdentity}) also in this case.

A suggestive approach \cite{CugliandoloKurchanPeliti} 
for justifying (\ref{SlopesIdentity}) consists in regarding 
$T_{\rm eff}(t,t_w)$ as the temperature measured by a thermometer coupled
to a particular observable of the system. It is quite natural 
to think that the result of this measure should not depend upon the 
observable. In aging systems with more than just one time sector,
this approach is not consistent unless the following identity holds:
\begin{eqnarray}
\frac{\chi_i(t,t_w)}{1-C_i(t,t_w)} = \frac{\chi_j(t,t_w)}{1-C_j(t,t_w)}\equiv
\frac{1}{T_{\rm movie}(t,t_w)}\, .
\label{NewTemperature}
\end{eqnarray}
Proving this conclusion will be the object of the next Section.
The new effective temperature $T_{\rm movie}(t,t_w)$ is in fact the
one measured
by a particular class of thermometers which we shall denote as ``sharp''.
It is a weighted average of the effective temperatures
(in the sense of (\ref{SlopesIdentity})) corresponding to different time
sectors.

Let us notice that Eq. (\ref{NewTemperature}) is remarkably
well verified in our discontinuous spin-glass model, cf. Fig.
\ref{Movieh01T04}, although it breaks down for $(t/t_w)\gg 1$.
In Sec. \ref{CoarseningSection} we demonstrated that it does not hold
for coarsening systems, and in fact a different relation is
true in this case, cf. Eq. (\ref{FittingTemperature}). Finally, 
we were not able to reach any definite conclusion 
for the Viana-Bray model of Sec. \ref{ContinuousSection}.
%
%
\section{Thermometric interpretation}
\label{ThermoApp}

One of the most striking (and unexpected) observations we made on our 
numerical results is summarized in Eq. (\ref{NewTemperature}). In this 
Section we shall try to connect this empiric observation to 
thermometric considerations.
This type of approach was pioneered in Ref. \cite{CugliandoloKurchanPeliti}
with the aim of interpreting the OFDR in terms of effective temperatures.
Here we will show that, in a system with more than a single time scale, 
this interpretation  is well founded if and only if
the ``movie'' temperature (\ref{NewTemperature})
can be defined.  In order to obtain this result, we shall carefully
reconsider the arguments of Refs. 
\cite{CugliandoloKurchanPeliti,ExartierPeliti,GarrigaRitort}.

According to Ref. \cite{CugliandoloKurchanPeliti}, the temperature of
a out-of-equilibrium system can be measured by weakly coupling it to 
a ``thermometer'', i.e. to a physical device which can be 
equilibrated at a tunable temperature $T_{\rm th}=1/\beta_{\rm th}$. 
The temperature of the system is defined as the value of $T_{\rm th}$ such that
the heat flow between it and the thermometer vanishes. The details
of the thermometer are immaterial in the weak-coupling limit. What matters
are the correlation and response functions of the 
thermometer\footnote{More precisely: $C_{\rm th}(t,t_w)$ and 
$R_{\rm th}(t,t_w)$
are the correlation and response functions for the observable of the 
thermometer which is coupled to the system.}
$C_{\rm th}(t,t_w)= C_{\rm th}(t-t_w)$ and 
$R_{\rm th}(t,t_w)= R_{\rm th}(t-t_w)$,
which are assumed to satisfy FDT: 
$R_{\rm th}(\tau) = -\beta_{\rm th} \partial_{\tau}C_{\rm th}(\tau)$.

In the spirit of our work, we shall couple the thermometer to a single 
spin variable $\sigma_i$  between times  $0$ and $t$,
and average over many thermal histories.
The measured temperature $\beta_{\rm th}$ is given by
\cite{CugliandoloKurchanPeliti,ExartierPeliti} 
\begin{eqnarray}
\beta_{\rm th}\int_0^t\!\! dt_w\,\, R_{\rm th}(t-t_w)
\partial_{t_w}C_i(t,t_w) = \int_0^t\!\! dt_w\,\, R_{\rm th}(t-t_w)
(-\chi_i'[C_i(t,t_w)])\partial_{t_w}C_i(t,t_w)\, ,
\label{TemperatureDefinition}
\end{eqnarray}
where we assumed the general OFDR (\ref{OFDR}) in its 
integrated form: $\chi_i(t,t_w) = \chi_i[C_i(t,t_w)]$, and denoted
by a prime the derivative of $\chi_i[\cdot]$ with respect to its argument.
Notice that {\it a priori} the measured temperature depends upon 
$i$ and $t$, for a given thermometer.

It is convenient to change variables from $t_w$ to $q\equiv C_i(t,t_w)$. 
This relation can be inverted by defining the time
scale $\tau_i(t;q)$ as follows: 
\begin{eqnarray}
C_i(t,t-\tau_i(t;q))=q\, .\label{DefinitionTimeScale}
\end{eqnarray}
Using these definitions in Eq. (\ref{TemperatureDefinition}),
we get
\begin{eqnarray}
\beta_{\rm th}\int_{q_{\rm min}}^1\!\!\!dq\,\, R_{\rm th}(\tau_i(t;q)) =
\int_{q_{\rm min}}^1\!\!\!dq\,\, R_{\rm th}(\tau_i(t;q)) (-\chi_i'[q])\, ,
\label{MeasuredTemperature_Q}
\end{eqnarray}
where $q_{\rm min}\equiv C_i(t,0)$. As $t\to\infty$, we have
$q_{\rm min}\to 0$. In the same limit 
$\tau_i(t;q)\to \tau^{\rm eq}_i(q)$ if $q>\qei$, while 
$\tau_i(t;q)\to \infty$ if $q<\qei$.

In order to measure temperatures on long time scales, we need a
thermometer with an adjustable time scale. Mathematically speaking, we take
$R_{\rm th}(\tau) = \Rt(\tau/\tau_{\rm th})$,
and use $\tau_{\rm th}$ to select the time scale. The precise 
form of $\Rt(x)$ is not very important. We shall assume that 
$\Rt(x)\approx 1$ for $x\ll 1$ and $\Rt(x)\approx 0$
for  $x\gg 1$. A simple example is $\Rt(x) =\theta(x)\, e^{-x}$. 
Some of the relations we will derive simplify 
if $\Rt(x)\approx \theta(x)\theta(x_*-x)$. We will call such a thermometer 
``sharp''.

We have two type of choices for the thermometer time scale $\tau_{\rm th}$:
\begin{enumerate}
\item We may take a ``fast'' thermometer, whose relaxation is much faster than
the structural rearrangements of the system. Equivalently: we look at our 
thermometer after a time $t\gg \tau_{\rm th}$. Mathematically this 
corresponds to taking the limit $t\to\infty$ with $\tau_{\rm th}$ fixed.
The result of such a measure is (for large times $t$) the bath temperature.
\item We may use a ``slow'' thermometer, with a relaxation time which is of 
the same order of the time needed for a structural change in the system.
This corresponds to taking the
limits $t\to\infty$, $\tau_{\rm th}\to\infty$ at the same time. If the system
ages, the outcome of such a measure will depend upon the precise way
these limits are taken.
\end{enumerate}
Let us consider separately the two cases.
%
%
\subsection{``Fast'' thermometer}
In this case we have, as $t\to\infty$,
\begin{eqnarray}
R_{\rm th}(\tau_i(t;q))\to\left\{\begin{array}{cl}
F_i(q) = \Rt(\tau_i^{\rm eq}(q)/\tau_{\rm th}) & \mbox{for $q>\qei$,}\\
0 & \mbox{for $q<\qei$,}
\end{array}\right.
\end{eqnarray}
with $F_i(\qei)=0$ and  $F_i(1)=1$. Inserting into Eq. 
(\ref{MeasuredTemperature_Q}) we get
\begin{eqnarray}
\beta_{\rm th}\int_{\qei}^1\!\!dq\, F_i(q) =
\int_{\qei}^1\!\!dq\, F_i(q) (-\chi_i'[q])\, .
\end{eqnarray}
Assuming that in the ``quasi-equilibrium'' time sector (i.e. for
$C_i(t,t_w)>\qei$) the system satisfies FDT, we can use
$\chi_i'[q] = -\beta$, which yields $\beta_{\rm th} = \beta$, as expected.
%
%
\subsection{``Slow'' thermometer}
Here we shall assume that the system has $p$ discrete correlation scales
in the aging regime \cite{CugliandoloKurchanWeak}. 
The generalization to a continuous set of correlation 
scales is straightforward.
To each scale $\alpha\in\{1,\dots,p\}$
we associates a time-scaling function $h_{\alpha}(t)$. 
As discussed in Sec. \ref{DiscussionSection},
$h_{\alpha}(t)$ is site-independent.

In order to probe the correlation scale $\alpha$, we tune the 
thermometer time scale with the function $\tau_{{\rm th},A}(t)$.
This function is defined by imposing
\begin{eqnarray}
\frac{h_{\alpha}(t)}{h_{\alpha}(t-\tau_{{\rm th},A}(t))} = A\, , 
\end{eqnarray}
for some fixed number $A>1$. 

Within the scale $\alpha$, we have 
$q_{\alpha+1}^{(i)}<C_i(t,t')<q_{\alpha}^{(i)}$.
It is easy to show that
\begin{eqnarray}
\lim_{t \to\infty}R_{\rm th}(\tau_i(t;q)/\tau_{{\rm th},A}(t)) =
F_{i,\alpha}(q)\, ,
\end{eqnarray}
 with
$F_{i,\alpha}(q)=0$ for $q<q_{\alpha+1}^{(i)}$,
$F_{i,\alpha}(q)=1$ for $q>q_{\alpha}^{(i)}$ and $F_{i,\alpha}(q)$
increasing in $(q_{\alpha+1}^{(i)},q_{\alpha}^{(i)})$.
Integrating by parts Eq. (\ref{MeasuredTemperature_Q}), we get
\begin{eqnarray}
\beta^{(i)}_{\rm th}\int_0^1 \!dq\, F_{i,\alpha}'(q) (1-q)=
\int_0^1 \!dq\, F_{i,\alpha}'(q) \,\chi_i(q)\, ,
\label{FinalTemperature}
\end{eqnarray}
which is our final expression for the temperature measured on the spin $i$
(here we emphasized the dependence of $\beta_{\rm th}$ upon the site).

Notice that the support of $F_{i,\alpha}'(q)$ is contained in the
interval $(q_{\alpha+1}^{(i)},q_{\alpha}^{(i)})$. The expression
(\ref{FinalTemperature}) simplifies in two cases: $(i)$ 
if the $\alpha$-th correlation scale is ``small''
$q_{\alpha+1}^{(i)}\approx q_{\alpha}^{(i)}\approx q^{(i)}_*$
(and, in particular, when there is a continuous set of scales); $(ii)$ 
if the thermometer is ``sharp'' in the sense defined at the beginning of
this Section, and, therefore, $F_{i,\alpha}'(q)$ is 
strongly peaked around some $q^{(i)}_*$. 
In both cases we have
\begin{eqnarray}
\beta^{(i)}_{\rm th} \approx \frac{\chi_i(q^{(i)}_*)}{1-q^{(i)}_*}\, .
\label{Temperature_Approx}
\end{eqnarray}
Let us now imagine to couple two copies of the same thermometer 
to two different sites $i$ and $j$. We shall measure two temperatures 
$\beta^{(i)}_{\rm th}\approx \chi_i(q^{(i)}_*)/(1-q^{(i)}_*)$ and
$\beta^{(j)}_{\rm th}\approx \chi_j(q^{(j)}_*)/(1-q^{(j)}_*)$,
with $q^{(i)}_* = f_{ij}(q^{(j)}_*)$.
These two temperatures coincide,  
$\beta^{(i)}_{\rm th}\approx \beta^{(j)}_{\rm th}$, {\it only if} 
Eq. (\ref{NewTemperature}) is satisfied.

The conclusion of the arguments presented so far is that
the condition (\ref{NewTemperature}) is {\it necessary} if we want a
given thermometer to measure the same temperature on any 
two spins of the system.
Moreover this condition is {\it sufficient} for the special class of ``sharp''
thermometers. 
In the last part of this Section we will show that 
the condition (\ref{NewTemperature}) is indeed {\it sufficient} 
for any thermometer, once (\ref{SlopesIdentity}) is assumed. 
%
%
\subsection{Thermometric equivalence of different sites}

We want to prove that Eqs. (\ref{NewTemperature}) and (\ref{SlopesIdentity})
imply the identity of thermometric temperatures on the sites $i$ and $j$
for any given thermometer. Let us stress that the measured temperature
may, eventually, depend upon the thermometer. The essential
ingredient for the ``small intropy production''
scenario of Ref. \cite{CugliandoloKurchanJapan}
to be applicable, is that the result should not depend upon the site.

Notice that from the definition (\ref{DefinitionTimeScale}),
it follows that the time scales defined on different sites are related as 
follows
\begin{eqnarray}
\tau_i(t;f_{ij}(q)) = \tau_j(t;q)\, ,
\end{eqnarray}
whence we easily derive the identity $F_{i,\alpha}(f_{ij}(q)) 
= F_{j,\alpha}(q)$. By the change of variables $q\to f_{ij}(q)$ we get, 
from Eq. (\ref{FinalTemperature})
\begin{eqnarray}
\beta^{(i)}_{\rm th}\int_{q_{\alpha+1}^{(j)}}^{q_{\alpha}^{(j)}} 
\!dq\,\, F_{j,\alpha}'(q) (1-f_{ij}(q))=
\int_{q_{\alpha+1}^{(j)}}^{q_{\alpha}^{(j)}} 
\!dq\,\, F_{j,\alpha}'(q) \,\chi_i(f_{ij}(q))\, ,
\end{eqnarray}
where we specified the range of $q$ such that $F_{j,\alpha}'(q)$ is
(possibly) nonzero. If use Eq. (\ref{NewTemperature}) to
connect the responses on different sites, we obtain
\begin{eqnarray}
\beta^{(i)}_{\rm th}\int_{q_{\alpha+1}^{(j)}}^{q_{\alpha}^{(j)}} 
\!dq\,\, F_{j,\alpha}'(q) (1-q)
\left[\frac{1-f_{ij}(q)}{1-q}\right]=
\int_{q_{\alpha+1}^{(j)}}^{q_{\alpha}^{(j)}} 
\!dq\,\, F_{j,\alpha}'(q) \,\chi_j(q)\left[\frac{1-f_{ij}(q)}{1-q}\right]\, .
\end{eqnarray}
The factors $(1-f_{ij}(q))/(1-q)$ prevent us 
from concluding that $\beta^{(i)}_{\rm th} = \beta^{(j)}_{\rm th}$ 
with no further assumption. 
Let us assume Eq. (\ref{SlopesIdentity}),
and that  $\chi_i'[q]$ stays constant for $q_{\alpha+1}^{(i)}<q
<q_{\alpha}^{(i)}$.
It follows that, within the scale $\alpha$, 
$f_{ij}(q) = 1-f^{0}_{\alpha,ij}(1-q)$,
$f^{0}_{\alpha,ij}$ being a constant. This implies $\beta^{(i)}_{\rm th} 
= \beta^{(j)}_{\rm th}$ for any thermometer. 
%
%
\section*{Acnowledgements}

This work received financial support from the ESF programme SPHINX and
the EEC network DYGLAGEMEM.
%
%
\appendix

\section{Large-$n$ calculations}
\label{AppLargeN}

In this Appendix we sketch the large-$n$ calculations whose results were 
presented in Sec. \ref{LargeNSection}.

%
%
\subsection{Statics}
\label{AppLargeN_Statics}

The trick for solving the periodic model of  Sec. \ref{LargeNSection}
is quite standard \cite{LargeNReview}. 
We define the $n\cdot V$-components vector $\psg_x$
which contains $n$ components for each type of spin:
\begin{eqnarray}
\psi^{a,u}_x = \phi^{a}_{x\cdot l+u}\,\, ,\;\; u\in\Lambda\,,\,\,x\in
{\mathbb Z}^d \, ,
\end{eqnarray}
where $x\cdot l =\sum_{\mu=1}^d x_{\mu}l_{\mu}$ and $\Lambda$ is the elementary
cell. In this basis the Hamiltonian reads
\begin{eqnarray}
H(\psg) = -\sum_{x,\mu}\psg_x\cdot \hat{\mathbb K}^{(\mu)}\psg_{x+\mu}-
\frac{1}{2}\sum_x \psg_x\cdot \hat{\mathbb L}\psg_{x}\, ,
\end{eqnarray}
where $\hat{\mathbb K}_{au,bv}^{(\mu)} = \delta_{ab}{\mathbb K}_{u,v}^{(\mu)}$,
$\hat{\mathbb L}_{au,bv} = \delta_{ab}{\mathbb L}_{u,v}$ and 
${\mathbb K}_{u,v} = J_{u,\hat{\mu}l_{\mu}+v}$, ${\mathbb L}_{u,v}=J_{u,v}$.

The equilibrium correlation functions are computed by standard methods:
\begin{eqnarray}
\<\psi_x^{a,u}\> & = &\delta^{a,1}M_u\, ,\\
\<\psi_x^{a,u}\psi_y^{b,v}\>_c & = & T \int_{BZ}\!\frac{dp}{(2\pi)^d}\,
[\M^{-1}_*(p)]_{u,v}\, e^{ip(x-y)}\, ,
\end{eqnarray}
where the $V\times V$ matrix $\M_*(p)$ is given by
\begin{eqnarray}
\M_*^{uv}(p) = -\sum_{\mu=1}^d \left[{\mathbb K}^{(\mu)}_{uv}
e^{ip_{\mu}}+ {\mathbb K}^{(\mu)}_{vu}e^{-ip_{\mu}}\right]-
{\mathbb L}_{uv}+\zeta^u_*\delta_{uv}\, .\label{MassMatrix}
\end{eqnarray}
The $V$ Lagrange multipliers $\zeta^u_*$ and the $V$ magnetizations $M_u$ must
be computed from the set of $2V$ equations given below:
\begin{eqnarray}
\sum_{v\in \Lambda} \M_*^{uv}(0) M_v = 0\, ,\label{LargeNSaddle1}\\
1 = M_u^2 + T\int_{BZ}\!\frac{dp}{(2\pi)^d}\,
[{\mathbb M}^{-1}_*(p)]_{uu}\, .\label{LargeNSaddle2}
\end{eqnarray}
These equations have two type of solutions: at high temperature
$M_u = 0$ and the matrix $\M_*(0)$ has rank $V$; at low temperature
$M_u >0$ and the matrix $\M_*(0)$ has one vanishing eigenvalue. 

In the following Section we shall treat the dynamics of this model. 
Remarkably all the complication produced by inhomogeneous couplings 
affects the aging dynamics only through the values of the local magnetizations
$\{ M_u\}$, the critical temperature $T_c$ and one more constant, $\Delta$,
which we are going to define. 
Consider the lowest lying eigenvalue $\lambda^0(p)$ of the matrix $\M_*(p)$.
As $p\to 0$ the corresponding eigenvector  coincide with $M_v$
and $\lambda^0(p)\to 0$. We then define
\begin{eqnarray}
\Delta = {\rm Det}\left[\left.\frac{\partial^2\lambda^0(p)}
{\partial p_{\mu}\partial p_{\nu}}\right|_{p=0}\right]\, .
\end{eqnarray}
All these quantities can be easily computed once the solution 
to Eqs. (\ref{LargeNSaddle1}), (\ref{LargeNSaddle2}) is known.
%
%
\subsection{Dynamics}
\label{AppLargeN_Dynamics}

The Langevin equation (\ref{Langevin}) is easily solved by 
defining the new order parameter $\psg_x$ as in the previous Section,
going to Fourier space: 
\begin{eqnarray} 
\partial\psi^{a,u}(p) = -\sum_{v\in\Lambda} \M^{uv}(p,t)\psi^{a,u}(p)+\eta^{a,u}(p,t)\, .
\end{eqnarray}
The ``mass'' matrix $\M^{uv}(p,t)$ is given by the expression 
(\ref{MassMatrix}) with the Lagrange multipliers $\zeta^u_*$ replaced 
by their time-dependent version $\zeta^u(t)$. 
Of course $\lim_{t\to\infty}\zeta^u(t) = \zeta^u_*$. 

The correlation and response functions for the field $\psg_x$ become
$V\times V$ matrices. Their diagonal elements are the on-site 
correlation and response functions of the field $\phg$. Standard 
manipulations yield:
\begin{eqnarray}
C(t,t_w) & = & \int_{BZ}\!\frac{d^dp}{(2\pi)^d} \, 
U(p;t)\left[1+2T\int_0^{t_w}\!\!ds\, U(p;s)^{-1}U(-p;s)^{-1}\right]U(-p;t)\\
R(t,t_w) & = & \int_{BZ}\!\frac{d^dp}{(2\pi)^d} \, U(p;t)U(p;t)^{-1}\, .
\end{eqnarray}
The matrix $U(p;t)$ satisfy the differential equation
\begin{eqnarray}
\partial_tU(p;t) = -\M(p;t)U(p;t)\, , \;\;\; U(p;0) = {\mathbb I}\, ,
\end{eqnarray}
and the Lagrange multipliers must satisfy the self-consistency conditions 
$C_{uu}(t,t) = 1$.

One can find the following asymptotic behavior for $U(p;t)$:
\begin{eqnarray}
U(p;t) = At^{d/4}(1+\gamma t^{-d/2+1}+\dots)\, e^{-\M_*(p)t}\, . 
\end{eqnarray}
The constants $A$ and $\gamma$ are simple numbers given below:
\begin{eqnarray}
A & = & \sqrt{\frac{\sum_{u\in\Lambda} M_u^2}{1+T/T_*}}
(8\pi)^{d/4}\Delta^{1/4}\, .\label{Aconstant}\\
\gamma & = & -\frac{T}{(\sum_{u\in\Lambda} M_u^2)(8\pi)^{d/2}\Delta^{1/2}}
\frac{\Gamma(1-d/2)^2}{\Gamma(2-d)}\, .\\
\end{eqnarray}
The constant $T_*$ appearing in Eq. (\ref{Aconstant}) is defined as
follows
\begin{eqnarray}
\frac{1}{2T_*}\equiv \int_0^{\infty}\!\!\!dt\,\,\, \hs\cdot U(0;t)^{-2}\hs\, ,
\label{UselessConstant}
\end{eqnarray}
where $\hs$ is the $V$-dimensional unit vector parallel to the vector
of the magnetizations: $\hat{\sigma}_u = M_u/(\sum_u
M_u^2)^{1/2}$. The expression (\ref{UselessConstant}) is quite hard to
evaluate, but this is not a problem, because $T_*$ cancels out in all
physical quantities.

Using the results listed above one can recover the general form 
(\ref{CoarseningCorrelation}) and the expressions 
(\ref{CoarseningLeading})-(\ref{CoarseningCorrection2}).
The universal functions which determine the domain wall contributions 
are given below for general dimension $2< d <4$ (we recall
that in the $n\to\infty$ limit the model is well defined in non-integer 
dimensions):
\begin{eqnarray}
{\cal F}_C(\lambda;d) & = & 
\frac{\Gamma(1-d/2)^2}{\Gamma(2-d)}\left(\frac{\lambda+\lambda^{-1}}{2}\right)^{-d/2}(1+\lambda^{2-d})-\nonumber\\
&&-\lambda^{d/2}\left(\frac{1+\lambda^2}{2}
\right)^{-d+1}\, 
\int_0^{2(1+\lambda^2)^{-1}}\!\!\!\!dx\,x^{-d/2}(1-x)^{-d/2}\, ,
\label{CcorrApp}\\
{\cal F}_{\chi}(\lambda;d)& = & \lambda^{-d+2}\int_1^{\lambda}\!dx\,
x^{d-3}(x^{d/2}-1)(x^2-1)^{-d/2}\, .\label{ChicorrApp}
\end{eqnarray}
The integral in Eq. (\ref{CcorrApp}) diverges for $d>2$: it is understood 
that it has to be analytically continued \cite{Gelfand}
 from $d<2$ to obtain the correct result. 

It can be useful to consider the asymptotic behavior of the expressions 
(\ref{CcorrApp}) and (\ref{ChicorrApp}). As $\lambda\to\infty$
(i.e.$t\gg t_w$) we have 
\begin{eqnarray}
{\cal F}_C(\lambda;d) & = & 2^{d/2-1}\lambda^{-d/2}
\left\{\left[\frac{\Gamma(1-d/2)^2}{\Gamma(2-d)}+\frac{4}{d-2}\right]+
\frac{\Gamma(1-d/2)^2}{\Gamma(2-d)}\lambda^{-d+2}+O(\lambda^{-2})
\right\}\, ,\\
{\cal F}_{\chi}(\lambda;d) & = & \lambda^{-d+2}\left\{\left[\frac{\Gamma(1-d/2)
\Gamma(1-d/4)}{2\Gamma(2-3d/4)}+\frac{1}{d-2}\right]-
\frac{2}{4-d}\lambda^{-2+d/2}+O(\lambda^{-3+d/2})
\right\}\, .\nonumber\\
\end{eqnarray}
As already remarked in Sec. \ref{LargeNSection} 
both functions vanish in the  $\lambda\to\infty$ limit.

When $\lambda\to 1$ one gets
\begin{eqnarray}
{\cal F}_C(\lambda;d) & = & \frac{d}{2-d}(\lambda-1)^{-d/2+1}
\left[1+O(\lambda-1)\right]\, ,\\
{\cal F}_{\chi}(\lambda;d) & = &\frac{2^{-d/2}d}{4-d}(\lambda-1)^{2-d/2}
\left[1+O(\lambda-1)\right]\, .
\end{eqnarray}
%
%
%

\end{document}